\documentstyle[12pt]{article}

\newcommand{\be}{\begin{equation}}
\newcommand{\ee}{\end{equation}}
\newcommand{\ba}{\begin{eqnarray}}
\newcommand{\ea}{\end{eqnarray}}

\newcommand{\al}{\alpha}

\newcommand{\Tr}{\rm Tr}
\newcommand{\tr}{\rm tr}

\begin{document}
\hoffset=-.4truein\voffset=-0.5truein
\setlength{\textheight}{8.5 in}
\begin{titlepage}
\begin{center}

{\large \bf Punctures and $p$-spin curves from matrix models}
\vskip .6 in
\begin{center}
{\bf E. Br\'ezin$^{a)}$}{\it and} {\bf S. Hikami$^{b)}$}
\end{center}
\vskip 5mm
\begin{center}
{$^{a)}$ Laboratoire de Physique de l'Ecole normale sup\'erieure, ENS,
Universit\'e PSL, CNRS, Sorbonne Universit\'e, Universit\'e de Paris, F-75005 Paris, e-mail: brezin@lpt.ens.fr}\\
{$^{b)}$ Okinawa Institute of Science and Technology Graduate University, 1919-1 Tancha, Okinawa 904-0495, Japan.
e-mail: hikami@oist.jp
} \\
\end{center}

\vskip 3mm       
{\bf Abstract}                  
\end{center}
  
This article investigates  the intersection numbers of the moduli space of $p$-spin curves with the help of matrix models. The explicit integral representations that are derived for the generating functions of these intersection numbers  exhibit $p$ Stokes domains, labelled by a "spin"-component $l$  taking values  $l = -1, 0,1,2,...,p-2$.  Earlier studies concerned  integer values of $p$, but the present formalism allows one to extend our study to half-integer or negative values of $p$,  which turn out to describe new types of punctures or marked points on the Riemann surface. They fall into two classes : Ramond $(l=-1)$, absent for positive integer $p$,  and Neveu-Schwarz $(l\ne -1)$.  The  intersection numbers of both  types are computed from the integral representation of the $n$-point correlation functions in a large $N$ scaling limit. We also consider a supersymmetric extension of the random matrix formalism to show that it leads naturally to an additional logarithmic potential. Open boundaries on the surface, or admixtures of R and NS punctures, may be handled by this extension.

  \end{titlepage}
\vskip 3mm

 \section{Introduction}
Although the connection  between  matrix models 
and the geometry of surfaces is by now an old subject, originating with the work of 't Hooft \cite{'t} for  large N gauge theories, new features have been revealed recently involving super-Riemann surfaces with punctures or boundaries,  related to matrix models with various types of symmetries \cite{Stanford}.
 
 In this work we return to the use of  Gaussian 
  matrix models with an external matrix source which, according to the nature of the source, covers  a large spectrum of applications. For instance we have considered earlier an external source which generates a gapped density of eigenvalues, and a critical point at the closing gap situation \cite{BrezinHikami90}.   In the course of this work this closing gap situation corresponds to the $p=3$ model defined below. As a quantum topological field theory, we have shown earlier that it provides a generating function of the intersection numbers  in the moduli space of Riemann surface of $p$-spin curves\cite{BrezinHikami01,BrezinHikami02,BrezinHikami03,BrezinHikami5,BrezinHikami2}.  This $p$ spin curve is described by an Hermitian matrix model involving the $(p+1)$-th power of  matrix, with an external source  tuned appropriately \cite{BrezinHikami2}, in a large matrix-size scaling limit. The model  is thus a  generalization of the Kontsevich Airy matrix model ($p=2$) \cite{Kontsevich}. 
 
 In this article, we consider the extension  to  half-integer spins such,  $p=\frac{1}{2},\frac{3}{2}, \cdots$, and 
to negative integers  for which new features appear \cite{BrezinHikami2, Semenov,Troost1}. Positive integer $p$  corresponds to a  compact coset space $SU(2)/U(1) $ with level $k=p-2$ of  the WZW model, and  negative $p$  to the non-compact, hyperbolic space,  $SL(2,R)/U(1)$, as   explained by Witten \cite{WittenP,Witten40}.  

 When $p= -1$, the intersection numbers reduce to  the Euler characteristics
of  orbifolds \cite{BrezinHikami03,Harer,Penner}, and when $p=-2$ the model is equivalent to a unitary matrix model with a $U(N)$ gauge field \cite{BrezinHikami2, Semenov,GrossWitten, BrezinGross, Gross, BrezinHikami70}. (A simple proof of the equivalence between a unitary matrix model with  the $p=-2$ model may be found  in the
appendix C of \cite{BrezinHikami4},  where it was shown that they share the same  equations of motion).

The  case of half-integer $p$, obtained by  continuation from  integer  $p$,   provides new features. The fractional level case corresponds to a fractional level $k$ of the WZW model, which is a non-unitary conformal field theory  \cite{Kac,Gaberdiel,Lesage}.  It is also related to the super ghost $\beta$-$\gamma$ system, within a superconformal field theory 
\cite{Nekrasov}. Another supersymmetric derivation of the intersection numbers of $p$-spin curves, which is different from the present article, may be found in \cite{Li,Verlinde}.
The case  half-integer $p$ is  also interesting as it realizes a Chern-Simons theory coupled to a Majorana fermion.
 It has been discussed earlier that the presence of fermions leads to the shift of the level  $k \to k - \frac{1}{2}$,  see for instance \cite{Gomis}.
The spin $ p= \frac{1}{2}$ case may thus be interpreted as $p= 1-\frac{1}{2}$, where the first term  $p=1$ comes from a simple  Gaussian case. Similarly   $p= - \frac{1}{2}$  may be interpreted as  a pure  fermion contribution, i.e. $p= 0 - \frac{1}{2}$.  For  both cases, 
we show that the intersection numbers involve Ramond punctures.

Ramond (R) and Neveu-Schwarz (NS) punctures come from the study of string theory over super-Riemann surfaces \cite{WittenR,WittenS}. 
Two types of nodal points can be defined on a  Riemann surface. If a line bundle at such a point is locally free (i.e. there is no
orbifold structure or in our approach no Stokes lines)
 it is called R-type ; otherwise (non-trivial orbifold structure) it belongs to the  NS- type.
 In this article, we
use these terminologies for the two distinct types of  punctures (or marked points). The contour  integrals  which describe the intersection numbers with these punctures turn out to be 
different.  For the NS- type, the component $l$ of  spin $p$ takes one of the values ($0,1,...,p-2$), and  the R-type $l$ corresponds to the single  value 
$l=-1$ \cite{Jarvis, Mochizuki}, which distinguishes the orbifold structures. We consider $s$ punctures on  a Riemann surface, and from a dimensional counting, a selection rule (Riemann-Roch, denoted RR) relates th
genus $g$, the spin $p$ and  the indices $n_i,l_i$ $(i=1,...,s)$.
It is well known that there are no contributions of R-type for  positive integer $p$ \cite{BrezinHikami5,Fan} as it will be shown in (\ref{onepoint}) of section 2. 

We  have also  found an R-contribution in the presence of an additional  logarithmic potential \cite{BrezinHikami5,BrezinHikami4}.
The R-type is  related to  the coefficients $m$ of  the logarithmic potential of the matrix model, in which the power  $m^b$, represents the number of  open boundaries of the Riemann surface.  One aim of this article is to discuss more systematically the appearance of R-type punctures on the basis of the results of 
\cite{BrezinHikami5,BrezinHikami4}, in which  open intersection numbers have already been considered.

 The calculations  of intersection numbers of R-type reduce to  computing residues from an integral representation.
  When $p \in Z+\frac{1}{2}$ (half-integer),  the moduli space becomes a spin moduli space, and it is related  for the lowest fractions to a Dirac spin ($p= \frac{1}{2}$)  or a 
  Rarita-Schwinger operator ($p= \frac{3}{2}$) \cite{Seeley,Rarita}.

When $p=-1/2$, all intersection numbers with one marked point  vanish  for a genus  $g > 1$, since   all the  intersection numbers of $p$-spin curves involve a  $(2p+1)$-factor. This behavior was noticed earlier from  explicit expressions   previously obtained up to genus 9 \cite{BrezinHikami2}. 
We   prove  here that for $p=-1/2$ the intersection numbers vanish to all orders in the genus $(g > 1)$.

 We have also considered the formalism of random supermatrices in an external source introduced in a previous article \cite{BrezinHikami3}. There the tuning of the source in a scaling limit generates a matrix model with and additional logarithmic potential term, a generalized Kontsevich-Penner model. The open intersection numbers had been computed earlier \cite{BrezinHikami4}. In this case, the R-type $(l=-1)$ does appear in the intersection numbers $\tau_{n,-1}$. (In \cite{BrezinHikami4}, it was denoted as $\tau_{n-\frac{1}{2}}$ ). These  R-type punctures are related to the odd or even character of the power of  $m$, in the coefficient of the logarithmic potential. For $p=\pm \frac{1}{2}$, an R-puncture is interpreted as a fermion on the boundaries; an odd number of R-punctures appears with an odd number of boundaries, and  an even number of R-punctures   only with  even powers of $m$ . Even  powers $b$  for $m^b$  correspond to an even number of boundaries. 
 
 We also discuss  a supersymmetric model with admixtures of  positive and negative  powers $p$ for the  matrices, in addition to a logarithmic term.  The negative integer $p$, and half-integer $p$ , matrix models were recently proposed in the context of  irregular conformal blocks \cite{Gaiotto,NishinakaRim} :  we briefly mention this correspondence in the section 5. The super-matrix formulation may shed light on the study of  the extension of $A_n$ singularities to other ADE cases \cite{Fan}, but such investigations are left for a future study. 

  This article consists of  the following sections.
In Section 2, we briefly review the formulae for  the $n$-point functions in  presence of an external source, and  how its tuning can generate a  $p$-spin curve  and provide  the intersection numbers.
In Section 3, we consider the one-point function. The extension to half-integer $p$  yields punctures of R-type. 
In Section 4, we study the two point  correlation function for positive integer spin $p$ and half-integer $p=\frac{1}{2}$ and obtain the intersection numbers. 
 In Section 5, we discuss  super-matrix models. Within such models, the two point correlation function is computed  for $p=\frac{1}{2}$(Ramond)  and $p=2$ (Neveu-Schwarz). 
  In Section 6, open intersection numbers are discussed through a logarithmic potential. 
  The Section 7 is devoted to a summary and discussions.
 In an appendix,  we consider the  intersection numbers for logarithmic potentials from  Virasoro equations. 
 
  {\section{ Intersection numbers for $p$-spin curves }}
  \vskip 2mm

  \vskip 2mm
  The  $p$-spin matrix model is a generalization of Kontsevich Airy matrix model \cite {Kontsevich} defined by 
\be\label{matrix}
Z = \int dB e^{-\frac{1}{p+1}{\rm Tr} B^{p+1} + {\rm Tr} B \Lambda}
\ee
where $B$ is a $k\times k$ Hermitian matrix, and $\Lambda$  a fixed $k\times k$ Hermitian matrix.  For the $p=2$ model,  Kontsevich  proved that $\log Z$, expanded in products of  inverse powers of the ${\rm Tr} (1/\Lambda^n)$, was a generating function of the intersection numbers  of the moduli of curves on a Riemann surface. The aim of this article is to derive those expansions for arbitrary $p$,  a priori a positive integer, but continued also to fractional or negative values.  The method consists of using two exact results a) an $N-k$  duality between $k$ point functions of characteristic polynomials  for $N\times N$ Hermitian matrices in an external matrix source b) explicit formulae for $k$ point functions for a Gaussian model in an external matrix source. These two results are summarized below. 

 We consider in the following  the average of a product of characteristic polynomials of the Hermitian $N\times N$ matrix $M$ with an external source $A$  defined by
   \ba\label{duality1a}
  F_k (\lambda_1,...,\lambda_k) &=& \frac{1}{Z_N}< \prod_{\alpha=1}^k {{\rm det}(\lambda_\alpha\cdot I - M)} >_{A,M}
  \nonumber\\
  &=& \frac{1}{Z_N}\int dM \prod_{\alpha=1}^k {\rm det}(\lambda_\alpha\cdot I - M) e^{- \frac{N}{2}{\rm tr} M^2 + N{\rm tr} M A}
  \ea
  where one averages over the $N\times N$ matrices $M$ ; $A$ is a given Hermitian matrix, whose eigenvalues are $(a_1,...,a_N)$.  $I$ is the  identity matrix and $Z_N$ is the normalization constant of the probability measure for $A=0$.   
  The duality formula derived in \cite{BrezinHikami01,BrezinHikami30} gives another matrix integral for the same average
  \be\label{duality2a}
F_k (\lambda_1,...,\lambda_k) = \frac{1}{Z_k}\int dB \prod_{j=1}^N {\rm det}(a_j - iB) e^{-\frac{N}{2}{\rm Tr} (B-i\Lambda)^2}
\ee
where $B$ is a  $k\times k$ Hermitian matrix and  $\Lambda= {\rm diag}(\lambda_1,...,\lambda_k)$ a source matrix; (we use the notation 'Tr' for traces of $k\times k$ matrices and 'tr' for those  $N\times N$). The normalization constant is  $Z_k = \int dB {\rm exp}(-\frac{N}{2}{\rm Tr} B^2)$. The proof is given in the appendix B of \cite{BrezinHikami01}.

Let us now  show that the freedom provided by the $N$ eigenvalues of the source $A$ may be used to tune the dual $B$ model (\ref{duality2a}) to  the p-spin matrix model (\ref{matrix}). For that purpose it is sufficient to specialize those formulae  to an external matrix source A possessing  $(p-1)$ distinct eigenvalues $A=(a_1,..,a_1,....,a_{p-1},....,a_{p-1})$. For simplicity we have assumed that  every distinct eigenvalue is degenerate $N/(p-1)$ times.
 Furthermore we now fix the eigenvalues of A  by the conditions
 \ba\label{pspin}
&&\sum_{i=1}^{p-1}\frac{1}{a_i^2} = p-1, \hskip 3mm \sum_{i=1}^{p-1}\frac{1}{a_i^m} = 0 \hskip 3mm (m=3,4,...,p)\nonumber\\
&&\sum_{i=1}^{p-1}\frac{1}{a_i^{p+1}}\ne 0.
\ea
 For example  consider the case $p=3$. Choosing $a_1=+1$, $a_2=-1$, one has $1/a_1^2+ 1/a_2^2 = 2$, $1/a_1^3+1/a_2^3 = 0$, $1/a_1^4+1/a_2^4 \neq 0$  which satisfy the conditions (\ref{pspin}).  \\
Dealing with the B-side of the duality (\ref {duality2a}) we consider  the expansion of 
 $\prod_{j=1}^N {\rm det}(a_j - iB)$ in inverse powers of the $a_j$ :
 $$ \prod_{j=1}^N {\rm det}(1- iB/a_j) = e^{-\frac{N}{p-1} \sum_{l=1}^{\infty}{\frac {1}{l}}{\rm Tr}(iB)^l \sum_{j=1}^{p-1} 1/a_j^l } $$
For definiteness consider  the $p=3$ case ; this expansion yields
  \be \label{determinant}  \prod_{j=1}^N {\rm det}(1- iB/a_j) = e^{ N {\rm Tr} (\frac{1}{2} B^2 - \frac{1}{4} B^4 +\cdots)}\ee
The $B^2$ terms in the exponential of (\ref{duality2a}) cancels  with the one in (\ref{determinant}) and one is left with  $ e^{N{\rm \Tr}( i\Lambda B - \frac{1}{4} B^4 + \cdots)}$. This is precisely the critical gap closing model studied in detail in \cite{BrezinHikami90}.
  Following Kontsevich \cite{Kontsevich} we want to expand the $B$-integral (more precisely the free energy $\log Z$) in powers of $ {\rm Tr} (1/ \Lambda ^n)$. To this effect we take a scaling limit in which the eigenvalues of $\Lambda$ are of order $ N^{-3/4}$ and the matrix elements of $B$ of order $N^{-1/4}$. In this scaling limit all the higher 
 terms of the form $N {\rm Tr} B^m, m>4$, in (\ref{determinant})  are negligible. 
 
For general $p$, following the conditions (\ref{pspin}),  the scaling limit  is given by $\Lambda \sim N^{-\frac{p}{p+1}}$ and $B\sim N^{-\frac{1}{p+1}}$ and the expansion  of (\ref{determinant}) stops with the highest term $N{\rm Tr} B^{p+1}$.  The two remaining terms $N{\rm Tr} \Lambda B$ and $N{\rm Tr} B^{p+1}$ are both of order one. So in the scaling limit the  $B$-side of the duality (\ref{duality1a}, \ref{duality2a}) is  indeed a generalized Kontsevich model (\ref {matrix}). 

It remains to compute the intersection numbers from the expansion of $F_k(\lambda_1, \cdots, \lambda_k)$ in powers of the 
\be \bar{t}_n = {\rm Tr} \frac{1}{ \Lambda^{n}} .\ee
i.e. as the coefficients of  ${\rm Tr } \frac {1}{\Lambda ^{n_1} }{\rm Tr } \frac {1}{\Lambda ^{n_2} }\cdots $.   

Now we appeal to the duality to compute them from the $M$-side  of the duality(\ref{duality1a}).  If we expand  $ \prod_{\alpha=1}^k {{\rm det}(\lambda_\alpha\cdot I - M)}$ in inverse powers of the $\lambda_i$'s, one can reconstruct these coefficients by combinatorics of the expectation values $< {\rm tr } M^{q_1}  {\rm tr } M^{q_2} \cdots >$ , given the relation
\ba <\prod_{a=1}^k \det (1-  \frac{M}{\lambda_a})> &=&< e^{\sum_{a=1}^k {\rm tr} \log( 1-M/\lambda_a) }>  \nonumber\\ & = &< e^{- \sum _{n=1}^{\infty}\frac{ \bar t_n {\rm tr} M^n}{n} }> \ea
where the bracket stands for an expectation value with the weight (\ref{duality1a}).  Therefore if the logarithm of the partition function (\ref{matrix}) is expanded in terms of the $ \bar t_{n_1} \bar t_{n_2} \cdots $ , the coefficients, i.e. the intersection numbers, are given by the connected expectation values $ <{\rm tr} M^{n_1}{\rm tr} M^{n_2}\cdots >_c $. 

  If the expression of $<{\rm tr}M^{n_1}\cdots {\rm tr}M^{n_q} >$ as functions of the  $a_j$'s is not simple, we know from our previous work  an exact expression for the n-point correlation functions 
\be  U(\sigma_1,....,\sigma_n) = <{\rm tr}e^{\sigma_1 M} \cdots {\rm tr}e^{\sigma_n M} > \ee
as function of the eigenvalues $a_j$.  These correlation functions are generating functions of the expectation values  $<{\rm tr}M^{n_1}\cdots {\rm tr}M^{n_q} >$. One  can then  reconstruct the expansion  of $F_k$ in powers of the $1/\lambda_i$'s from the expansions of the $U$'s in powers of  the $\sigma$'s. 

Indeed for arbitrary eigenvalues $a_j$ of the  source matrix $A$, we have found in \cite{BrezinHikami02} that
 \be \label{new}
U(\sigma) = <{\rm tr} e^{\sigma M} >=  \frac{1}{N\sigma}e^{\frac{N}{2}\sigma^2}\oint \frac{du}{2i \pi} \prod_{j=1}^N ( 1- \frac{\sigma}{a_j-u})e^{N\sigma u}
\ee
where the contour in the u-plane encloses the eigenvalues $a_j$.
Similarly the n-point correlation function $U(\sigma_1,...,\sigma_n)$ are given by contour integrals over n complex variables $\sigma_i$,
\ba\label{npoint}
&&U(\sigma_1,....,\sigma_n) = <{\rm tr}e^{\sigma_1 M} \cdots {\rm tr}e^{\sigma_n M} >\nonumber\\
\nonumber\\
&&=e^{\sum_{1}^n \sigma_i^2}\oint \prod_{i=1}^n \frac{du_i}{2i \pi}\ e^{\sum_1^n u_i \sigma_i} \prod_{\alpha=1}^N \prod_{i=1}^n
(1- \frac{\sigma_i}{a_\alpha - u_i})\  {\rm det}\frac{1}{u_i-u_j + \sigma_i}
\ea

The contours are taken  around the poles $u_i= a_\alpha$, not around the poles which come from the determinant.

We now specialize those formulae  to the  above external matrix source A possessing  $(p-1)$ distinct eigenvalues $A=(a_1,..,a_1,....,a_{p-1},....,a_{p-1})$ satisfying the constraints (\ref{pspin}). Then we expand, for reasons specified below, 
$$ \prod_{j=1}^N ( 1- \frac{\sigma}{a_j-u}) = e^{\frac{N}{p-1} \sum_1^{p-1} \log {(1-\frac {\sigma}{ a_j-u})}}$$
in inverse powers of the $1/a_j$, taking $\sigma_j$ and $u$ as small and same order.
For instance if $p=3$, with the  conditions  (\ref{pspin}), the  term $e^{N\sigma u}$ in (\ref{new}) cancels and, the leading term in the exponent is $e^{{N/4}[u^{4}-(u+\sigma)^{4}]}$. In  the large $N$ scaling range in which $N \sigma\Lambda \sim N^0$, i.e. $\sigma \sim N^{-1/4}$, we can neglect the powers in $u$ and $\sigma$ higher than $4$.  

For arbitrary $p$, with this choice of the matrix source A,  in the scaling range under consideration  $\Lambda \sim N^{-\frac{p}{p+1}}$ , $\sigma$ and $u$ of order $N^{-1/(p+1)}$ ,  we obtain  a generating function for the one-point intersection numbers of $p$-spin curves (i.e. the coefficients of single trace operators ${\rm Tr} (1/ \Lambda^n)$ from the integral  \ref{duality2a}),
\be\label{pintegral}
U(\sigma) = \frac{1}{N\sigma}{\rm exp}[-\frac{N}{p-1}\sigma (\sum_{j=1}^{p-1}\frac{1}{a_j})] \int \frac{du}{2i\pi} e^{{C}[u^{p+1}-(u+\sigma)^{p+1}]}
\ee
with $C= \frac{N}{p^2-1}\sum_{i=1}^{p-1} \frac{1}{a_i^{p+1}}$.   If we choose further the condition $\sum \frac{1}{a_j}=0$, the exponential factor in the front of the integral can be omitted. 

 With the conditions (\ref{pspin}), the average of the characteristic polynomials for the  matrix $B$ of (\ref{duality2a}) turns into the $p$-th generalized Kontsevich model  
(\ref{matrix}). In this matrix model (\ref{matrix}), we consider the eigenvalues $\lambda_j$ of the source  matrix $\Lambda$ as large, and
the intersection numbers are obtained from a $1/\Lambda$ expansion. In the dual formulation  the large $\Lambda$ expansion corresponds to an expansion of  $U(\sigma)$ for small $\sigma$. 
                 
In the large $N$ limit we obtain, with this choice of the matrix source A, a generating function for the intersection numbers of $p$-spin curves  \cite{BrezinHikami02},
\be\label{pintegral1}
U(\sigma) =\frac{1}{\sigma} \int \frac{du}{2i\pi} e^{{C}[u^{p+1}-(u+\sigma)^{p+1}]}
\ee
or, after translation, 
\be\label{pintegral2}
U(\sigma) = \frac{1}{\sigma}\int \frac{du}{2i\pi} e^{{C}[(u-\sigma/2)^{p+1}-(u+\sigma/2)^{p+1}]}
\ee
with $C= \frac{N}{p^2-1}\sum_{i=1}^{p-1} \frac{1}{a_i^{p+1}}$.  Since we have expanded the integrand   of (\ref{new}) in the scaling limit $u\sim\sigma\sim N^{- 1/(p+1)}$ the contours for the  re-scaled $u_i$'s  are now around infinity. For this  $p$-spin model, the  asymptotic expansion in $\sigma$  is governed by which of $p$ Stokes regions one is considering  in the complex $u$-plane. The asymptotic behavior of the integrand in (\ref{pintegral2}) being an exponential of $\sigma u^p$  threre are $p$ sectors given by the p-th roots of unity, labelled by an index from 0 to (p-1) or, in accordance with the mathematical literature, a "spin"-index $l=-1,0, \cdots,p-2$. 

Let us first consider the case $p=3$ for which (\ref{pintegral2}) leads to  an Airy function ${\rm A_i}(\zeta)$ (as in \cite{BrezinHikami01})
\be\label{Ai}
U(\sigma) = \frac{1}{(N\sigma)^{4/3} 3^{1/3}}{\rm A_i}(\zeta)
\ee
with $\zeta =-N^{2/3}4^{-1}3^{-1/3}\sigma^{8/3}$. 
The  Airy function $A_i(x)$ is defined by
\be\label{Airy}
A_i((3a)^{-\frac{1}{3}}x) = \frac{(3a)^{1/3}}{\pi} \int_0^\infty dt\  {\rm cos}(at^3+ xt).
\ee
The standard expansion of the Airy function for small $x$ consists of  two distinct series, and not three in spite of the three Stokes regions, 
\ba\label{Airy2}
U(\sigma)&=& \frac{1}{(N\sigma)^{4/3}3^{1/3}}[{\rm A_i}(0)(1+ \frac{1}{3!}\zeta^3 + \frac{1\cdot 4}{6!}\zeta^6+ \frac{1\cdot4\cdot7}{9!}\zeta^9+ \cdots)\nonumber\\
&+& {\rm A_i}'(0) (\zeta + \frac{2}{4!} \zeta^4 + \frac{2\cdot 5}{7!}\zeta^7 + \frac{2\cdot 5\cdot 8}{10!}\zeta^{10} + \cdots)]
\ea
where the first series corresponds to the spin-index $l=1$ with $\sigma^{m+\frac{2}{3}}$, and the second series to a spin-index $l=0$ with $\sigma^{m+\frac{1}{3}}$. 
The overall factors are $Ai(0)= \frac{1}{2\pi 3^{1/3}}\Gamma(\frac{1}{3})$ and $Ai'(0) = -\frac{1}{2\pi}\Gamma(\frac{2}{3})$.  

A priori a third Stokes region with spin-index $l=-1$ could have been present, which would correspond to an asymptotic expansion of the form $\zeta^{2+3m}$ (m=0,1,2,...), leading  to $\sigma^{\frac{8}{3}(2+ 3m)-\frac{4}{3}}= \sigma^{4+8m}$, but such "Ramond" contribution is absent from the final asymptotics.  

Similarly for arbitrary $p$, there are $p$ Stokes region labelled by the  spin index $l$. The asymptotic expansion  is generated by the integral
\ba\label{expansion}
&&U(\sigma)=\frac{1}{N\sigma}\int \frac{du}{2i\pi} {\rm exp}[- c\sigma u^p] \times \nonumber\\
&&  {\rm exp}[ - c(\frac{p(p-1)}{3! 4} \sigma^3 u^{p-2} + \frac{p(p-1)(p-2)(p-3)}{5! 4^2} \sigma^5 u^{p-4}+\cdots)] 
\ea
where  the constant $c$  is related  to $C$ in (\ref{pintegral2}) : $c=(p+1)C$ , namely
\be \label{constant} c = \frac{N}{p-1} \sum_{i=1}^{p-1} \frac{1}{a_i^{p+1}} \ee
a number proportional to N. Expanding in powers of $\sigma$ the second exponential and changing the integration variable $u \to (c\sigma)^{-1/p} u^{1/p} $,  with a phase which depends of the Stokes sector under consideration, one obtains for the $l$-th sector
\be\label{onepoint}
U_l(\sigma)=  \frac{1}{N}\sum_n<\tau_{n,l}>\  \frac{1}{n\pi}\Gamma(1 - \frac{1}{p} - \frac{l}{p}) c^{\frac{n+(l+1)/p}{p+1}}p^{ 1 + \frac{ pn+l+1}{p+1}}\sigma^{n +(l+1)/p}\ee
in which the spin index $l$  is a fixed number, taking  one of the values  $-1$ to $p-2$. The Gamma function in (\ref{onepoint}) and the power of $p$  have  been extracted to match the conventional normalization of the  intersection numbers $<\tau_{n,l}>$. 

The result is presented in (\ref{onepoint}) after expansion of the second exponential in (\ref{expansion}).  However it is interesting to return to the meaning of this expansion in terms of the $1/N$ expansion. The  coefficient $c$ is proportional to $N$ and the expansion  is performed for $c\sigma = O(1)$, i.e. $ \sigma = O(N^{-1}) $.Therefore collecting the powers of $1/N$ in the r.h.s. of (\ref{onepoint})  one finds that the n-th term is proportional to $1/N^{2g}$ with an index $g$ related to $n$ by
\be \label{nl}
(2g-1) (1 + \frac{1}{p})= n +\frac{1}{p}( l +1)
\ee
So, as in 't Hooft classical result, $g$ is indeed the genus of the Riemann surface under consideration.  Finally we may present the result (\ref{onepoint}) in the more transparent form
\be \label{onepoint2} U_l(\sigma) = \sum_g <\tau_{n,l}>_g \frac{1}{n\pi}\Gamma(1 - \frac{1}{p} - \frac{l}{p}) c^{\frac{2g-1}{p}}p^g \sigma^{(2g-1)(1+ \frac{1}{p})} \ee

 This generalizes the  asymptotic expansion discussed above for the  Airy case (p=3), with $l=1, 2$. 
  Again, like in the Airy case for $p=3$, the $l=-1$ term which would violate (\ref{nl}) with integer $n$ and $g$, is missing : all intersection
numbers $<\tau_{n,l}>$ are of  NS-type. 

It is interesting to note that (\ref{nl}) agrees in the case under consideration here with the Riemann-Roch relation (RR) which,  for $s$-marked points
(punctures), reads
\be\label{RR}
3g -3 + s =\sum_{i=1}^s n_i + (g-1)(1 - \frac{2}{p}) + \frac{1}{p}\sum_{i=1}^s l_i
\ee
or equivalently, 
\be
(2g -1)(1 + \frac{1}{p}) = \sum_{i=1}^s (n_i + \frac{1}{p}l_i) + (1 - s + \frac{1}{p})
\ee

For one marked point ($s=1$), this gives, as was derived in (\ref{nl})
\be
(2g-1) (1 + \frac{1}{p})= n +\frac{1}{p}( l +1)
\ee
This shows  the consistency of the matrix model results with the basic geometrical results derived by Witten \cite{WittenP}. The same approach in which one considers several marked points allows one to show that the matrix model is consistent with this RR geometrical rule.  Furthermore we shall argue later that the matrix model provides a generalization of this RR relation for half-interger values of $p$. 

Finally let us quote the simplest results that one derives from (\ref{onepoint}) 
 \ba\label{inters}
&&<\tau_{1,0}>_{g=1} = \frac{p-1}{24},\nonumber\\  &&<\tau_{n,l}>_{g=2} = \frac{(1+ 2p)(p-1)(p-3)}{p\cdot 5!\cdot 4^2\cdot 3}\frac{\Gamma(1-\frac{3}{p})}{\Gamma(1-\frac{1+l}{p})}\nonumber\\
&&<\tau_{n,l}>_{g=3} =\frac{(1+ 2p)(p-1)(p-5)(8p^2-13 p -13)}{p^2\cdot 7! 4^3 3^2} \frac{\Gamma(1 -\frac{5}{p})}{\Gamma(1- \frac{1+l}{p})}
\ea
where $n$ and $l$ have to  satisfy the condition  (\ref{nl}) which relates $n$ and $g$ at fixed $l$ ; otherwise the intersection numbers vanish. Up to genus $g=9$, explicit values of the intersection numbers had been obtained in \cite{BrezinHikami2}. Note the factors $(p-1)$ and $(2p+1)$ in the intersection numbers for genus $g$ on which we return  in Section 3 and show that for $g\ge 2$, these two factors are always present . 
In conclusion the index $l$ takes only the $(p-1)$ values $0,1,...,p-2$, which correspond to  NS-punctures.  The R -punctures  which could have occurred  for  $l=-1$ are absent.

\vskip 2mm
{\section{One point function for  integer  and half-integer  $p$ }}
\vskip 3mm
Let us first repeat what was done in the previous section in a slightly different setting, more adapted to extensions to half-integer or negative values of $p$. 
 We start as earlier from the representation of the one-point function $ <{\rm tr} e^{\sigma M} >$ in the scaling limit, given by (\ref{pintegral2}) :  
\be\label{zintegral}
U(\sigma) = \frac{1}{\sigma}\int \frac{du}{2i \pi} e^{C ((u- \frac{1}{2}\sigma)^{p+1} - (u + \frac{1}{2}\sigma)^{p+1})}.
\ee

We now change  variable 
\be\label{zy}
u = \frac{i}{2} ( y^2- y^{-2} ), \hskip 3mm du = i ( \frac{1}{y^3} + y ) dy
\ee
which will turn out to be better suited to discuss half-integer $p$.
The one point function $U(\sigma)$ of (\ref{zintegral}) becomes
\be\label{double}
U(\sigma)= \frac{i}{2}\oint \frac{dy}{2i\pi} (y+ \frac{1}{y^3})e^{C(\frac{i \sigma}{4})^{p+1} \frac{1}{y^{2p+2}}[(y^2-i)^{2p+2}-(y^2+i)^{2p+2}]}
\ee

The power $(2p+2)$  in (\ref{double}) is twice of  $(p+1)$ in (\ref{zintegral}) and it makes the continuation to half-integers $p$ easy. 
(This formulation for half-integer $p$ is related to the spin structure, in which the double covering for a half-integer spin is related to the spin moduli space
$\tilde M_g$ as shown in \cite{Seeley}.) The integral over $y$ may lead to both NS and R-punctures. As we will see  the NS type is described as  integrals along  cuts
in the  $t$-plane with $u= t^{\frac{1}{p}}$ in (\ref{onepoint}).   
 In addition the pole   at $y=0$, provides a non-vanishing residue, which corresponds to
a singularity of  R-type.

The one point correlation function is now written
\be\label{g}
U(\sigma) = \frac{i}{2}\int \frac{dy}{2i\pi} (\frac{1}{y^3}+ y) g(y)
\ee
with
\be\label{gy}
g(y) = {\rm exp} { \biggl[  C (\frac{i}{2})^{p+1}(\frac{\sigma}{2})^{p+1}\frac{1}{y^{2p+2}}\biggl((y^2-i)^{2p+2} -  (y^2+i)^{2p+2}\biggr)\biggr]}
\ee
Let us specify this integral for a few simple values of $p$.
\vskip 2mm
{\bf (i) positive \hskip1mm integer\hskip 2mm  p=1,2}
\ba\label{yy1}
g(y) &=& {\rm exp}[i \frac{1}{2} C\sigma^2 (y^2-y^{-2}) ] \hskip 10 mm (p=1)\nonumber\\
g(y) &=& {\rm exp}[ -C \sigma^{3} \frac{1}{16}( 3 y^4 - 10 + 3 \frac{1}{y^4})] \hskip 10mm (p=2)
\ea

{\bf (ii) negative \hskip 1mm integer\hskip 2mm p= -1, -2,-3}
\ba\label{yy2}
g(y) &=& {\rm exp} [ C \sigma^{-1} \frac{16 y^4}{(1+ y^4)}] \hskip 8mm (p=-2)\nonumber\\
g(y) &=& {\rm exp} [ - 128 i C \sigma^{-2}\frac{y^{10}-y^6}{(1+ y^4)^4}]\hskip 8mm (p=-3)
\ea
\hskip 8mm The case of $p=-1$ is obtained from a limit $p\to -1$ in (\ref{gy}), noting that  the constant c, given by (\ref{constant}), $c= \xi/(p+1)$, 
\be
g(y) = (\frac{y^2+i}{y^2-i})^{2\xi},\hskip 5mm (p=-1)
\ee

{\bf (iii) positive \hskip 1mm half \hskip1mm integer \hskip 2mm $p=\frac{1}{2},\frac{3}{2}$}
\ba\label{yy3}
g(y) &=& {\rm exp}[ C (i)^{1/2}\frac{1}{4} \sigma^{\frac{3}{2}}(3y - \frac{1}{y^3})] \hskip 10mm (p=\frac{1}{2})\nonumber\\
g(y) &=&  {\rm exp}[ C (i)^{3/2} \frac{1}{16} \sigma^{\frac{5}{2}}(5y^3  - \frac{10}{y} + \frac{1}{y^5})] \hskip 10mm (p=\frac{3}{2})
\ea

{\bf (iv) negative \hskip 1mm half \hskip1mm integer \hskip 2mm $p=-\frac{1}{2},-\frac{3}{2}$}
\ba\label{yy4}
g(y) &=& {\rm exp}[ - C (i)^{\frac{3}{2}} \frac{\sigma^{\frac{1}{2}}}{y}] \hskip 10mm (p=-\frac{1}{2})\nonumber\\
g(y) &=&  {\rm exp}[  4 C (i)^{\frac{1}{2}} \sigma^{-\frac{1}{2}}(\frac{y}{1+y^4})] \hskip 10mm (p=-\frac{3}{2})
\ea

\vskip 3mm
{\bf{(i) positive integer $p$ }}
\vskip 3mm
We now return to (\ref{g}) with positive integer $p$ to check that we recover in this new setting the results of section two. 
Let us start with the pure  Gaussian case ($p=1$) which reads  from (\ref{yy1}) and (\ref{g}),  with $c= \frac{i}{2} C$,
\be
U(\sigma) =\oint \frac{dy }{2i\pi}(\frac{1}{y^3}+ y) e^{c \sigma^2 (y^2-y^{-2})}
\ee

This is just  the contour integral of a total derivative and therefore it vanishes.  In particular there are no terms which are proportional to $\sigma^n$ in $U(\sigma)$, i.e. R-type.

To summarize in the trivial $p=1$ case $<\tau_{-2,-1}>_{g=0} = 0$. For  $g\ge 1$, there is no contribution, thus  all intersection numbers  vanish $<\tau_{n,l}>_g = 0$. We thus expect that the intersection numbers  should involve  a factor  ($p-1$). The presence of this factor had been already checked for the intersection numbers up to $g=9$ \cite{BrezinHikami2}. 

For   integer $p>1$, similarly the integral  (\ref{g}) over a circle vanishes. However again the presence of Stokes lines implies different asymptotic behaviors in different regions, as discussed at length in the previous section. In this formalism with integral (\ref{g}) over $y$, let us look for instance at the $p=2$ case with , from (\ref{yy1}), 
$$ g(y) =  {\rm exp}[ -a (  y^4  + \frac{1}{y^4})] $$
with $a = - 6 c (\frac{\sigma}{2})^3$
and  thus
\be U(\sigma) = \int {dy} (y + \frac{1}{y^3}) e^{-a(y^4+ \frac{1}{y^4})}\ee
Along the line at  ${\pi/4}$ this yields after change $ y^4\to t$ 
\ba && \int_0^{\infty}  dt ( t^{-\frac{1}{2}} + t^{-\frac{3}{2}}) e^{-a(t+ \frac{1}{t}}) \\ 
&=& \frac{1}{4 a^{\frac{1}{2}}}e^{\frac{10}{3}a}[ \Gamma(\frac{1}{2}) + a \Gamma(-\frac{1}{2}) - a^2 \Gamma(-\frac{1}{2}) - a^2 \Gamma(-\frac{3}{2}) + O(a^3)]
\ea
 This expansion contains , as seen earlier, only  NS  terms ,and it agrees with (\ref{onepoint}).

If we could not find   any R-puncture in $<\tau_{n,-1}>$  for positive integer $p$, it will be 
shown that it exists  for  (ii) negative values of $p$ and also for (iii),(iv) half-integer $p$. 
We consider half-integer
 $p=-\frac{1}{2}, -\frac{3}{2}, ...$ as well as negative integers $p=-1, -2, -3,...$ in the following.
\vskip 3mm

{\bf{(ii) negative integer $p$ }}
\vskip 3mm
Let us examine the consequence of the previous formulae if we extend them to negative values of $p$. First the connection between $n$ and $g$  established in (\ref{RR}) 
\be\label{selection1}
(2g-1) (1 + \frac{1}{p})= n +\frac{1}{p}( l +1)\ee
may be extended to negative integer $p$.  We have to specify the  spin-index $l$ which characterizes the integration sectors and takes now  $|p|$ different values,
\be\label{ll}
l= 0,-1,-2,-3,...,p+1 \hskip 3mm (p \le -1).
\ee
The R-type is still associated with  $l=-1$.  The choice of  negative values of $l$ in the selection rule (\ref{selection1}) for labelling the $|p|$ sectors, is reasonable 
since $\frac{l}{p}$ remains positive.


\vskip 3mm
{\bf{(ii-a) $p= -1$ case}}
\vskip 3mm
Since this case is a limit we  need to return here to the initial representation (\ref{pintegral2}) 
\be
U(\sigma) =\frac{1}{\sigma} \int \frac{du}{2i\pi} e^{{C}([u-\sigma/2)^{p+1}-(u+\sigma/2)^{p+1}]}
\ee
with $C= \frac{N}{p^2-1}\sum_{i=1}^{p-1} \frac{1}{a_i^{p+1}}$ to obtain the limit as $p$ goes to $-1$.
\be
U(\sigma) = \frac{1}{N}\int \frac{du}{2i \pi} \biggl( \frac{u-\frac{1}{2}}{u+\frac{1}{2}}\biggr)^N
\ee
 Writing 
$(u-\frac{1}{2})/(u+\frac{1}{2})= e^{-z}$ and $  u= (1+e^{-z})/(1-e^{-z})$,
we obtain
\be
U(\sigma) = -\frac{1}{N} \int_0^\infty \frac{dz}{2\pi} \frac{e^{-z}}{(1- e^{-z})^2} e^{- N z} \ee
This integral diverges linearly at the origin and a regularization is needed. As argued below this divergence is in fact linked to genus zero and after a zeta-regularization by the genus, the divergent genus zero term is discarded.  Then one finds
\ba
U(\sigma) &=& \int_0^\infty \frac{dz}{2\pi} (\sum B_n \frac{y^{n-1}}{n!}) e^{-Nz}\nonumber\\
&=& 1 - \frac{1}{2N} + \frac{1}{12 N^2} - \frac{1}{120 N^4} +\cdots
\ea
where the $B_n$ are Bernoulli numbers $ \frac{t}{1-e^{-t}}= \sum_0^{\infty} \frac{B_n}{n!} t^n $. For this  $p=-1$ case, there is no $\sigma$ dependence, and instead we have considered the dependence in the size  $N$ of the matrix. This leads to the Euler characteristics \cite{Harer, Penner, BrezinHikami03},
\be
\chi (\bar M_{g,1}) = \zeta(1 - 2g)= -  \frac{1}{2g}B_{2g}
\ee
Note that for  $p=-1$ we have no choice other than $l=0$ in the list (\ref{ll}), we have  the unique choice $l=0$ from (\ref{selection1}).\vskip 2mm
{\bf {(ii-b) p= - 2}}
\vskip 3mm
We are dealing with
\ba\label{minus2}
U(\sigma) &=& \int dy (y+ \frac{1}{y^3}) e^{- \frac{y^4}{\sigma(1+ y^4)}}  d y \nonumber\\
&=& \frac{1}{\sqrt{\sigma}}\int dx \frac{1}{x^2} e^{- \frac{4 x^2}{1 + \sigma x^2}}
\ea
where the equality is obtained first by noting the invariance of the integral under $y\to 1/y$  and then $y^2\to \sqrt{\sigma} x$.
This model is in fact equivalent to  a unitary matrix model as was shown  in \cite{BrezinHikami70}.  Then the expansion of the integral (\ref{minus2}) provides 
\be\label{unitary}
U(\sigma) = - \frac{1}{2 \sqrt{\pi \sigma}}( \frac{1}{8}\sigma + \frac{3^2 \sigma^2}{3!2^7} +
\frac{3^2\cdot 5^2}{5!2^9}\sigma^3 + \frac{3^2\cdot 5^2\cdot 7^2}{21\cdot 2^{18}}\sigma^4 +\cdots).
\ee

 It reproduces  the  results of the unitary matrix model. For the unitary matrix model, there are terms which come from the angular measure which yields a  logarithmic potential . The result of (\ref{minus2}) is obtained from the unitary matrix model with the coefficient  $m$ of logarithmic potential 
in the limit $m\to 0$ \cite{BrezinHikami70}.  We find $l=0$, and the expansion provides terms of the form
$\sigma^{g - \frac{1}{2}}$ for genus $g=1,2,3,...$. The intersection numbers are $<\tau_{n,0}>_g$, which are all of NS type.
There is no R-type  for the one point function.
The result  (\ref{unitary}) gives the intersection numbers,
\be
<\tau_{1,0}>_{g=1}= -\frac{1}{8}, \hskip 2mm <\tau_{2,0}>_{g=2} = \frac{1}{2^7}, \cdots\ee
which agrees with the general expression (\ref{inters}) for the intersection numbers of $p$ spin curves.
\vskip 3mm

{\bf{(ii-c) p= - 3}}
\vskip 3mm
The possible values of $l$ are
$l= -1, 0,1$.
The one point function $U(\sigma)$ becomes after the scaling $y = \sigma^{\frac{1}{3}}x$,
\be\label{-3p}
U(\sigma) = \int dx (\frac{\sigma^{-\frac{2}{3}}}{x^3} + \sigma^{\frac{2}{3}}x) e^{- a \frac{x^6(1-\sigma^{4/3}x^4)}{(1+ \sigma^{4/3}x^4)^4}}
\ee
From the selection rule  relative to $p=-3$, the possible  intersection numbers are
$
<\tau_{1,0}>_{g=1},<\tau_{2,-1}>_{g=2},<\tau_{4,1}>_{g=3},<\tau_{5,0}>_{g=4},...
$
in which only $l=-1$ is of R-type. 
We have from (\ref{-3p}),
\be
U(\sigma) = -\frac{1}{3}\sigma^{-\frac{2}{3}}+ \frac{5}{18}\sigma^{\frac{2}{3}}\Gamma(\frac{1}{3}) + 16 \sigma^2 + \cdots
\ee
and thus 
\be
<\tau_{0,1}>_{g=0} = -\frac{1}{3},\hskip 2mm <\tau_{1,0}>_{g=1}= \frac{5}{18}, \hskip 2mm <\tau_{2,-1}>_{g=2}= 16,\dots
\ee
the last one being of  R-type ($l=-1$)  \vskip 2mm

For  the case of negative even integers $(p=-2m)$, the selection rule (\ref{selection1}) forbids $l=-1$   since $\frac{(2g-1)}{p}$ is a half-integer. Therefore   we have no possibility of R-puncture when $p$ is an even negative integer,
whereas it does occur for  odd negative integers with $l=-1$.
 \vskip 3mm
{\bf{(iii) positive half-odd integer case $p\in \frac{1}{2}+ Z$}}
\vskip 3mm
The possible values of $l$ are now $2p$ instead of $p$  since we made the change of variables (\ref{zy}) which doubles the number of sectors. This leaves the possibilities of  
\be \label{positive half}  {l = - 1, - \frac{1}{2}, 0, \frac{1}{2},  ...., p-\frac{3}{2}} \ee.

\vskip 3mm
{\bf {(iii-a) p=$\frac{1}{2}$}}
\vskip 3mm
For $p=\frac{1}{2}$, the only  possible value is $l= - 1$, which is an  R-puncture.
We have from (\ref{yy3}), the one point function,
\be\label{l1/2}
U(\sigma) = \frac{i}{2} \oint \frac{dy}{2 i \pi} (\frac{1}{y^3}+y) e^{ c' \sigma^{\frac{3}{2}} (3y - \frac{1}{y^3})}
\ee
where $c'= \frac{1}{4} i^{1/2} C$.
The possible value of the spin component $l$ is only $l=-1$ from (\ref{positive half}).
 Indeed we have the selection rule for the intersection numbers in (\ref{nl}) $\sigma^{3(2g-1)}= \sigma^{n+ 2(l+1)}$ with $l= - 1$, which is integer power of $\sigma$. Hence the intersection numbers $<\tau_{n,l}>_g$ are with $n = 3+ 6m$ ($m=1,2,...)$. Non vanishing possible intersection numbers are
$
<\tau_{3, - 1}>_{g=1}, <\tau_{9, - 1}>_{g=2}, <\tau_{15, - 1}>_{g=3},....
$
These values may be obtained, after  a Taylor expansion of the second exponential term,
and picking up the residues at the $y=0$  pole.
This one point is  easily computed   in closed form by the same technique,

Thus we have
\be\label{p1/2sigma}
U(\sigma) = \frac{i}{2} \sum_{j=0}^\infty c'^{4j+2} \frac{3^{2+ 3j}}{j!(3j + 3)!} (-1)^j \sigma^{3+ 6 j}
\ee
 These terms are consistent with the selection rule, since the $g$ dependence in $\sigma$ is $\sigma^{3(2g-1)}$. 
 Using the expression of $U(\sigma)$ of (\ref{p1/2sigma}), we have an explicit result in closed form for  
 \be
 T(z) = \int_0^\infty U(\sigma) \sigma^4 e^{-\sigma^2 /z} d \sigma = \frac{9 }{4}i z^4 e^{{27 c'^4}{z^3}}
 \ee
 This equation shows the relation between $p=\frac{1}{2}$ and  the Airy distribution ($p=2$) with a logarithmic term.

\vskip 3mm
{\bf{(iii-b) p=$\frac{3}{2}$}}
\vskip 3mm
This spin $p=\frac{3}{2}$ may correspond to a Rarita-Schwinger operator \cite{Rarita, Seeley}. 

The possible values of $l$ are
$
l= - 1, - \frac{1}{2}, 0.
$
We have

\be\label{p3/2}
U(\sigma) = \oint \frac{dy}{2\pi i} (\frac{1}{y^3}+ y) e^{c (\frac{\sigma}{2})^{\frac{5}{2}}[ 5 y^3-\frac{10}{y}+\frac{1}{y^5}]}
\ee
This gives terms of order $\sigma^{\frac{5}{3}},\sigma^{5},\sigma^{\frac{25}{3}},...$, which are consistent with the selection rule for $p= \frac{3}{2}$.  
The expansion of $U(\sigma)$ gives the series of terms of  $<\tau_{n,l}>_g \sigma^{n+ \frac{2}{3}(l+1)}$, where the coefficients, i.e. intersection numbers, become  $<\tau_{-2,-\frac{1}{2}}>_{g=0}$, $<\tau_{1,0}>_{g=1},<\tau_{5, -1}>_{g=2},<\tau_{8,-\frac{1}{2}}>_{g=3}$. 
The term  $l= -1$ which gives an integer power of $\sigma$ is  of R-type.  At higher orders the R-type appears with $\sigma^5, \sigma^{15},\sigma^{25},...$.
These terms of integer power $\sigma^{5+10 m}, m=0,1,2,3,...$ are obtained from the contour integral along the circle $|y|=1$ where the pole at $y=0$ in (\ref{p3/2}) exisists, which are evaluated  as
\be
U_R(\sigma) = \frac{5}{2^5}c^2 \sigma^5 - \frac{7\cdot 5^4}{3\cdot 2^{16}}c^6 \sigma^{15} + \frac{79\cdot 11\cdot 5^7\cdot 3^2}{10!\cdot 2^{24}}c^{10}\sigma^{25} + \cdots
\ee
where the first term is $g=2$, and the second term is a $g=5$ contribution to a R-puncture. 

The NS-type intersection numbers are obtained from (\ref{p3/2}) in the sector defined by the replacement $ y^3 =  t$, and with $a= - c (\frac{\sigma}{2})^{5/2}$.
\ba
&&U_{NS}(\sigma) = \frac{1}{3}\int_0^\infty \frac{dt}{2 \pi} t^{-\frac{2}{3}} (t^{\frac{1}{3}}+ \frac{1}{t})e^{-a( 5 t - 10 t^{-1/3} + t^{-5/3})}\nonumber\\
&&= \frac{1}{3a}\int_0^\infty \frac{dt}{2\pi} a^{\frac{2}{3}} t^{-\frac{2}{3}}((\frac{t}{a})^{1/3} + \frac{a}{t})e^{- 5 t} \biggl ( 1 + \frac{10}{t^{1/3}}a^{4/3} + (\frac{50}{t^{2/3}}- \frac{1}{t^{5/3}})a^{8/3}+ \cdots\biggr)\nonumber\\
\ea
It provides a series expansion,
\ba\label{p3/2series}
&&U_{NS}(\sigma) = \frac{1}{15a} \int_0^\infty \frac{dx}{2\pi} (a^{1/3} (\frac{x}{5})^{-1/3} + a^{5/3} (\frac{ x}{5})^{-5/3}) e^{-x} ( 1 + \frac{10 a^{4/3}}{(\frac{x}{5})^{1/3}} + \cdots )
\nonumber\\
&&= \frac{5^{1/3}}{15}\Gamma(\frac{2}{3}) (- c (\frac{\sigma}{2}))^{-\frac{5}{3}}+ \frac{5^{5/3}}{15}\Gamma(-\frac{2}{3}) (- c(\frac{\sigma}{2}))^{\frac{5}{3}} + \cdots
\ea
This series gives terms with non-integer powers of $\sigma$ , thus of NS type, which are consistent with the selection rule for $p=\frac{3}{2}$. 
Indeed the series (\ref{p3/2series}) shows non-vanishing  intersection numbers  $<\tau_{-2,-\frac{1}{2}}>_{g=0}$, $<\tau_{1,0}>_{g=1}$, $<\tau_{5,-1}>_{g=2}$, ...., which were expected from the selection rule  $\sigma^{(2g-1)(1+ \frac{1}{p})}=\sigma^{n+ \frac{1}{p}(l+1)}$.

The case of $p=\frac{3}{2}$ is particularly interesting since the level $k$ of $su(2)$, ($k=p-2$), becomes $k=-\frac{1}{2}$.
The ${\mathcal N}=2$ super symmetric $ \widehat {su}(2)_{-\frac{1}{2}}$ WZW model is a non-unitary conformal field theory with  central charge $c= 3-\frac{6}{k+2}=-1$ \cite{Kac,Lesage},
and for the coset $\widehat{su}(2)_{-1/2}/u(1)$ WZW model, the central charge $c=-2$. The $\widehat{su}(2)$ means the affine super algebra \cite{Kac}. 
\vskip 3mm
{\bf{(iv) negative half-odd integer $p$}}
\vskip 3mm
For negative half-odd integers $p$, there are $2\vert p \vert$  allowed values of $l$, namely
$
l = -1, -\frac{1}{2}, 0, \frac{1}{2}, 1, \frac{3}{2},..., {|p|}-\frac{3}{2}.
$

\vskip 3mm
\noindent {\bf{ { (iv-a)} p= $- \frac{1}{2}$ case}}
\vskip 3mm

Since for $p=-\frac{1}{2}$ the factor $\frac{1}{p}(l+1)$ becomes
 an integer, the $\sigma$ dependence does not involve  fractional powers. Thus the $p=-\frac{1}{2}$ is nterpretated ias R-type punctures.

 $U(\sigma)$
has here the expression 
\be\label{UA}
U(\sigma) =   \int dy (\frac{1}{y^3}+ y) e^{c \sqrt{\sigma}\frac{1}{y}}
\ee

Expanding in $\sigma$ the contour integral, the only non-vanishing term  is 
   $\frac{c^2}{2}\sigma$. 
Since the intersection number $<\tau_{n,l}>$ is the coefficient of the term $\sigma^{n+ \frac{1}{p}(l+1)}$ (= $\sigma^{(2g-1)(1+\frac{1}{p})}$) in the expansion of $U(\sigma)$,
this term corresponds to   $n=1, l= - 1$, with $g=0$. This corresponds to an R-puncture ($l=- 1$) for $g=0$ with non-zero $<\tau_{1,- 1}>_{g=0}$. 

The first  term $\frac{1}{y^3}$ does not contribute to the pole at $y=0$. Given the essential singularity at the origin one has to be more careful. 

Indeed if one changes $y\to -\frac{1}{x}$,  the first term of the  integral becomes 
\be
\int_0^\infty dx x e^{-c\sqrt{\sigma} x} = \frac{1}{c^2\sigma}
\ee
which could not be found in a $\sigma$ expansion. This result  corresponds tp a term $\sigma^{n+ \frac{l+1}{p}}$ with $p=-\frac{1}{2}, n=-1,l= -1$. This leads to 
a non-zero intersection number $<\tau_{-1,-1}>_{g=1}$.
 The expansion has just two terms of order $\sigma$ and $\sigma^{-1}$ , and there are no more terms. This is consistent with the factor $(2p+1)$ in the intersection numbers for general $p$ (as shown in (\ref{onepoint}) \cite{BrezinHikami2}), which vanishes for $p=-\frac{1}{2}$.

\vskip 3mm

\noindent {\bf{(iv-b) p=  $ - \frac{3}{2}$ }}
\vskip 3mm
For $p=-\frac{3}{2}$,  there are $2\vert p\vert$ allowed values for  $l$  are
$
l= - 1, - \frac{1}{2}, 0
$
and the selection rule for $g,n,l$ reads 
\be\label{selection(-3/2)}
\sigma^{(2g-1)(1+ \frac{1}{p})}= \sigma^{(2g-1)\frac{1}{3}} = \sigma^{n- \frac{2}{3}(l+1)}
\ee
Therefore for $g=0$, it gives the power $\sigma^{-\frac{1}{3}}$, which leads to $n=0,l=-\frac{1}{2}$, i.e. $<\tau_{0,-\frac{1}{2}}>_{g=0}$.
For $g=1$, it becomes $\sigma^{\frac{1}{3}}$, and we obtain $n=1,l=0$, i.e. $<\tau_{1,0}>_{g=1}$.
For $g=2$, we $\sigma^{(2g-1)(1+\frac{1}{p})}= \sigma^{(2g-1)\frac{1}{3}}= \sigma$. Thus we have $n=1,l= -1$,
$<\tau_{1,-1}>_{g=2}$, an R-type case.

 For $g=3$, we have $\sigma^{(2g-1)\frac{1}{3}}= \sigma^{\frac{5}{3}}$,
then  $n=2,l=-\frac{1}{2}$, $<\tau_{2,-\frac{1}{2}}>_{g=3}$. 

For $g=4$, we have similarly $<\tau_{3,0}>_{g=4}$ with $\sigma^{\frac{7}{3}}$. For $g=5$, we have $\sigma^3$ and
$<\tau_{3,-1}>_{g=5}$. This is again an R-type.
Thus we find  R-punctures at   $g= 2+ 3m$, ($m=0,1,2,...$),  for which $U(\sigma)$ yields integer powers $\sigma^{2m+1}$, and the  intersection numbers are 
\be
<\tau_{2m+1,-1}>_{g=2+ 3m},\hskip 5mm (m=0,1,2,...)
\ee
We  can verify these results by the explicit calculation of  (\ref{g}).

We take $c 2^{\frac{1}{2}}\sigma^{-\frac{1}{2}} = \sigma'^{-\frac{1}{2}}$ for simplicity. The one point function is
\be\label{Ry}
U(\sigma)  = \int dy (\frac{1}{ y^3} +  y) {\rm exp}({- \sigma'^{-\frac{1}{2}}\frac{y}{1+  y^4}})
\ee
which gives an expansion in integer powers of $\sigma'$ if we rescale $y$ by $\sigma'^{1/2}y$. This leads to  R-intersection numbers with a $\sigma^{2m+1}$ ($m=0,1,2,...)$) dependence.

 When we change  $y \to  \sigma'^{-\frac{1}{6}} t^{-\frac{1}{3}} $ in (\ref{Ry}), we obtain
   \ba\label{Ny}
 U(\sigma) &=& - \frac{1}{3}\int dt ( \sigma'^{-\frac{1}{3}} t^{-\frac{5}{3}} + \sigma'^{\frac{1}{3}}t^{-\frac{1}{3}})
 {\rm exp}({- \frac{t}{1 + \sigma'^{\frac{2}{3}}t^{\frac{4}{3}}}})\nonumber\\
 &=& (-\frac{1}{3})[\sigma'^{-\frac{1}{3}}\Gamma(-\frac{2}{3})+ \frac{5}{2}\sigma'^{\frac{1}{3}}\Gamma(\frac{2}{3}) + 3 \sigma' + O(\sigma'^{\frac{5}{3}})]
\ea
 
 The expansion of this integral for small $\sigma'$ gives  terms of order $\sigma^{\frac{1}{3}(2g-1)}$, (g=0,1,2,...). This expansion is consistent with the result of the selection rule (\ref{selection(-3/2)}). Thus we have an R-puncture $l= -1$, with  integer powers of $\sigma$, for $g=2+ 3m$, ($m=0,2,3,...)$ . The dependence in $\sigma$ for the R-type reads $\sigma^{(2g-1)\frac{1}{3}}
 = \sigma^{( 2m+1)} $ with $g= 2+ 3m$, which is consistent with the R-calculation of the integral (\ref{Ry}).  We also find NS-types with fractional powers of $\sigma$ by (\ref{Ny}). \vskip 3mm

This analysis may be easily extended to  $p= -\frac{7}{2},- \frac{9}{2},  ....$, and we find both R ($l=-1$) and NS ($l\ne -1$) contributions.


Thus we have studied many cases provided by the representation (\ref{g}) of the one point function.
We have found R and  NS-punctures in those cases. The computation of the intersection numbers of 
R-punctures for $p= \frac{1}{2},\frac{3}{2}, ...$ results from a contour integral around the pole at $y=0$. For negative
integers and   negative half-integers, the intersection numbers are obtained by the same technique, and they yield both  R  and NS- punctures.

\vskip 3mm
 {\section{Two point correlation function $U(\sigma_1,\sigma_2)$ }}
\vskip 3mm

Higher correlation function, involving the intersection numbers which appear as coefficients of several $\bar t_n$  in the expansion of the free energy,  can be handled by the same technique.  Again one considers in the $1/\lambda_a$ expansion of $<\prod_{a=1}^K \det (1-  \frac{M}{\lambda_a})> $  the coefficients which involve $< \tr M^{n_1} \tr M^{n_2}\cdots>$. A generating function for these coefficients are the $U(\sigma_1, \cdots,\sigma_n)$. Let us consider 
the  two point function 
\be
U(\sigma_1,\sigma_2) = <{\rm tr} e^{\sigma_1 M}{\rm tr}e^{\sigma_2 M} >.
\ee
It is given  by the formula  (\ref{npoint}) for $n=2$,
\be\label{npoint1}
U(\sigma_1,\sigma_2) = 
e^{\sum_{1}^n \sigma_i^2}\oint \prod_{i=1}^2 \frac{du_i}{2i \pi}\ e^{\sum_1^n u_i \sigma_i} \prod_{\alpha=1}^N \prod_{i=1}^2
(1- \frac{\sigma_i}{a_\alpha - u_i})\  {\rm det}\frac{1}{u_i-u_j + \sigma_i}
\ee
The diagonal part of the determinant gives the disconnected part $<{\rm tr}e^{\sigma_1 M} ><{\rm tr}e^{\sigma_2 M} >$.  The off -diagonal term gives the connected correlator which, after the shift $u_1\to u_1 -\frac{1}{2}\sigma_1 , u_2\to u_2 -\frac{1}{2}\sigma_2$, deals with a product, which is conveniently replaced by an additional integration
\ba&& \frac{1}{u_1-u_2+ \frac{1}{2}(\sigma_1+\sigma_2)} \frac{1}{u_1-u_2-\frac{1}{2}(\sigma_1+\sigma_2)} \nonumber\\
& =&\frac{2}{\sigma_1+\sigma_2}\int_0^\infty
dx e^{- x(u_1-u_2)} {\rm sinh} \frac{x}{2}(\sigma_1+\sigma_2)
\ea

\vskip 2mm
{\bf Positive integer $p$ }
\vskip 2mm
We now return to the large-N scaling  \cite{BrezinHikami03}  which led to the $p$-spin formulae with $\sigma_i\sim N^{-1/(p+1)}$,  with the same scaling expansion of $ \prod_{\alpha=1}^N \prod_{i=1}^2
(1- \frac{\sigma_i}{a_\alpha - u_i})$  for the source specified by the conditions (\ref{pspin}).  This yields 
\ba\label{dx}
&&U_c(\sigma_1,\sigma_2) = \frac{2}{\sigma_1+\sigma_2}\frac{1}{(2i\pi)^2}\int_0^\infty dx \int du_1du_2 \hskip 2mm {\rm sinh}(\frac{1}{2}x (\sigma_1+\sigma_2)) e^{-x(u_1-u_2)}\nonumber\\
&&\times {\rm exp}[ -\frac{N}{p^2-1} c \sum_{i=1}^2((u_i+\frac{1}{2}\sigma_i)^{p+1}-(u_i-\frac{1}{2}\sigma_i)^{p+1} )]
\ea
The expression (\ref{dx}) has the form
 \ba U_c(\sigma_1,\sigma_2) &=& \frac{2}{\sigma_1+\sigma_2}\frac{1}{(2i\pi)^2}\int_0^\infty dx \int du_1du_2 \hskip 2mm {\rm sinh}(\frac{1}{2}x (\sigma_1+\sigma_2))\nonumber\\ && e^{-x(u_1-u_2)} G(u_1,\sigma_1) G(u_2, \sigma_2) \ea  
 with \be G(u, \sigma) = {\rm exp}[ -\frac{N}{p^2-1} c((u+\frac{1}{2}\sigma)^{p+1}-(u-\frac{1}{2}\sigma)^{p+1} )]. \ee These expressions may all be expanded in powers of $\sigma_1$ and $\sigma_2$ and provide 
a generating function of the two-point intersection numbers.
The lowest order in $\sigma_1$  is a term of order $\sigma_1^{\frac{1}{p}}\sigma_2^{2+ \frac{1}{p}}$ ;  in agreement with (\ref{RR}), the RR selection rule, this gives a non-zero $<\tau_{0,0}\tau_{2,0}>_{g=1}$ 
\be
<\tau_{0,0}\tau_{2,0}>_{g=1} = \frac{p-1}{24}
\ee
which is equal to  $<\tau_{1,0}>_{g=1}$ as implied by the  string equation, or lowest Virasoro constraint for the equations of motion, 
\be
<\tau_{0,0}  \tau_{n,l}>_g = <\tau_{n-1,l}>_{g}
\ee

When $p=2$, we find
\ba\label{symp=2}
&&U(\sigma_1,\sigma_2) = \frac{2}{(\sigma_1+\sigma_2)\sqrt{\sigma_2}}e^{\frac{1}{24}(\sigma_1^3+\sigma_2^3)}
\int_0^\infty dx \ {\rm sinh}(x \frac{\sqrt{\sigma_1}}{2}(\sigma_1+\sigma_2))e^{-\frac{\sigma_1+\sigma_2}{\sigma_2}x^2}\nonumber\\
&=& \frac{1}{\sigma_1+\sigma_2}e^{\frac{1}{24N'^2}(\sigma_1+\sigma_2)^3}\sum_{m=0}^\infty
\frac{(-1)^m}{m!(2m+1)}(\frac{\sigma_1\sigma_2(\sigma_1+\sigma_2)}{8})^m \sqrt{\sigma_1\sigma_2}\nonumber\\
&=& \sum_{n_1,n_2}<\tau_{n_1,0}\tau_{n_2,0}>_g \sigma_1^{n_1+\frac{1}{2}}\sigma_2^{n_2+\frac{1}{2}}
\ea
Thus we have obtained  explicit formulae for the intersection numbers $<\tau_{n_1,o}\tau_{n_2,0}>_g$ at arbitrary  genus $g$. The above formula for $p=2$ may be expressed in a symmetric way as
\be
U_c(\sigma_1,\sigma_2) = \frac{2}{(\sigma_1+\sigma_2)}e^{\frac{1}{24}(\sigma_1^3+\sigma_2^3)}
\int_0^\infty dx\ {\rm sinh}(x \frac{\sqrt{\sigma_1 \sigma_2}}{2}(\sigma_1+\sigma_2))e^{-(\sigma_1+\sigma_2) x^2}
\ee

This does not yield any integer power of the $\sigma_i$, since  ${\rm sinh}(x)$ is odd :  the powers of $\sigma_i$ ($i=1,2$) are all half-integers. Hence the two-puncture points are of NS- type, no R-puncture  for $p=2$.

 For $p=3$, similarly from
 the explicit expression for the  two-point function \cite{BrezinHikami66}, one does not find  R-punctures. Indeed
for $p=3$, we have \cite{BrezinHikami66},
\ba
U_c(\sigma_1,\sigma_2) &=& \frac{2}{(\sigma_1+\sigma_2)(3 \sigma_2)^{1/3}} \int_0^\infty dx \hskip 2mm {\rm sinh} (\frac{\sigma_1+\sigma_2}{2}(3 \sigma_1)^{1/3} x ) A_i ( x - \frac{1}{4\cdot 3^{1/3}}\sigma_1^{8/3})\nonumber\\
&\times& A_i(- (\frac{\sigma_1}{\sigma_2})^{1/3}x - \frac{1}{4\cdot 3^{1/3}}\sigma_2^{8/3})
\ea
where the Airy function $A_i(x)$ is defined by 
\be
A_i(x) = \int_{-\infty}^\infty \frac{du}{2\pi} e^{\frac{i}{3}u^3 + i u x}
\ee
The above formula may be written in a symmetric way as
\ba
&&U_c(\sigma_1,\sigma_2) = \frac{2}{(\sigma_1+\sigma_2)} \int_0^\infty dx \hskip 2mm {\rm sinh} (\frac{\sigma_1+\sigma_2}{2}(3 \sigma_1)^{1/3} (3\sigma_2)^{1/3} x ) \nonumber\\
&&\times A_i ( (3\sigma_2)^{1/3} x - \frac{1}{4\cdot 3^{1/3}}\sigma_1^{8/3})
 A_i(- (3 \sigma_1)^{1/3} x - \frac{1}{4\cdot 3^{1/3}}\sigma_2^{8/3})
\ea
Note that the Airy function $A_i(x)$ decays exponentially  for $x\to + \infty$, and oscillates when $x\to -\infty$. Therefore the integral over $x$ is finite.
The $p=3$ intersection numbers for two points are all of Neveu-Schwarz type.
More details for this $p=3$ case may be found in \cite{BrezinHikami66}.

For $p=4$, we have similarly 
\ba
U_c(\sigma_1,\sigma_2) &=& \frac{2}{(\sigma_1+\sigma_2)(4\sigma_2)^{1/4}}\int_0^\infty dx \int_0^\infty
dv_1 dv_2 \hskip 1mm {\rm sinh}(\frac{\sigma_1+\sigma_2}{2}(4 \sigma_1)^{1/4} x)\nonumber\\
&\times & e^{-\frac{\sigma_1^3}{2}(\frac{1}{4\sigma_1})^{1/2}v_1^2 - \frac{\sigma_2^3}{2}(\frac{1}{4\sigma_2})^{1/2} v_2^2}
e^{-\frac{1}{4}v_1^4+ x v_1 - \frac{1}{4}v_2^4-ax v_2}
\ea
where $a= (\sigma_1/\sigma_2)^{1/4}$. 
This may be written as
\ba
U_c(\sigma_1,\sigma_2) &=& \frac{2}{(\sigma_1+\sigma_2)}\int_0^\infty dx 
 \hskip 1mm {\rm sinh}(\frac{\sigma_1+\sigma_2}{2}(4 \sigma_1)^{1/4}(4\sigma_2)^{1/4} x)\nonumber\\
&\times & \phi_{+}(x)\phi_{-}(x)
\ea
where
\be
\phi_{+}(x) = \int dv e^{-\frac{1}{4}v^4 - \frac{1}{4}\sigma_1^{5/2}v^2 + (4\sigma_2)^{1/4} x v}
\ee
\be
\phi_{-}(x) = \int dv e^{-\frac{1}{4}v^4 - \frac{1}{4}\sigma_2^{5/2}v^2 - (4\sigma_1)^{1/4} x v}
\ee

The two-points intersection numbers may be obtained  for larger values of $p$  by this method \cite{BrezinHikami66}.

\vskip 2mm
{\bf  (ii) half-integer p  }
\vskip 2mm
Using the  expression (\ref{yy3}) of $g(y)$, 
the two point correlation function $U(\sigma_1,\sigma_2)$  is given by
\ba\label{twopoint}
U(\sigma_1,\sigma_2) &=& 
\oint \frac{dy_1dy_2}{(2\pi i)^2} (\frac{1}{y_1^3}+ y_1)(y_2+\frac{1}{y_2^3}) g(y_1)g(y_2)\nonumber\\
&& \frac{1}{(\frac{\sigma_1}{2}u_1-\frac{\sigma_2}{2}u_2)^2- \frac{1}{4}(\sigma_1+\sigma_2)^2}
\ea
where $u_i = \frac{i}{2}(y_i^2 - y_i^{-2})$and $g(y_i)$ is given by (\ref{yy3}) and (\ref{yy4}).

\vskip 2mm
{\bf (ii-a) p = $\frac{1}{2}$}
\vskip 2mm
The   $p=\frac{1}{2}$  case presents, as we have seen for the one point function in (\ref{p1/2sigma}), 
 a spin component $l= -1$). The two point function for $p=\frac{1}{2}$ is given  from (\ref{twopoint}) and, after the
shift $y_i\to \sigma_i^{-\frac{1}{2}}y_i$, $c'^2= (\frac{i}{2})^{p+1}  c^2$, it reads
\ba\label{p1/2 int}
U_c(\sigma_1,\sigma_2) &=&  4 \oint \frac{dy_1dy_2}{(2\pi i)^2} (y_1+ \frac{\sigma_1^2}{y_1^3})(
y_2+ \frac{\sigma_2^2}{y_2^3})e^{c' \sigma_1 (3  y_1- \frac{\sigma_1^2}{y_1^3}) + c' \sigma_2( 3  y_2 - \frac{\sigma_2^2}{y_2^3})}\nonumber\\
&&\times \frac{1}{(y_1^2-\frac{\sigma_1^2}{y_1^2}- y_2^2+ \frac{\sigma_2^2}{y_2^2})^2 + 4 ( \sigma_1+ \sigma_2)^2}
\ea

The selection rule (\ref{RR}), $3g - 3 + s = \sum n_i + (g-1)(1-\frac{2}{p}) + \frac{1}{p}\sum l_i$, becomes  with $s=2$ (two-point), $l_i= -1$,
\be\label{6g}
6g = n_1+n_2
\ee
Thus we expect the  series  $U(\sigma_1,\sigma_2)=
\sum_{g,n_1,n_2}  C_{g,n_1,n_2} c'^{4g} \sigma_1^{n_1}\sigma_2^{n_2}$ with integers $n_1$ and $n_2$, and coefficient $C_{g,n_1,n_2}$.

The two point function is expanded for small $\sigma_i$ as
\ba
U(\sigma_1,\sigma_2) &=& 4 \oint \frac{dy_1 dy_2}{(2i\pi)^2} (y_1+ \frac{\sigma_1^2}{y_1^3})(y_2 + \frac{\sigma_2^2}{y_2^3}) \frac{1}{(y_1^2-y_2^2)^2}\nonumber\\
&& \times \frac{1}{1- f} e^{c'\sigma_1(3y_1- \frac{\sigma_1^2}{y_1^3}) + c' \sigma_2 (3 y_2- \frac{\sigma_2^2}{y_2^3})}
\ea
where 
\be
f = \frac{2}{y_1^2-y_2^2}(\frac{\sigma_1^2}{y_1^2}- \frac{\sigma_2^2}{y_2^2}) - \frac{1}{(y_1^2-y_2^2)^2}((\frac{\sigma_1^2}{y_1^2}-\frac{\sigma_2^2}{y_2^2})^2 + 4 
(\sigma_1+\sigma_2)^2)
\ee
The factor $\frac{1}{1-f}$ is expanded as $\sum_{m=0}^\infty f^m$.
The two point function $U(\sigma_1,\sigma_2)$ is  expanded in the power of $\sigma_1,\sigma_2$ and $c'$, which becomes a series of   $c'^{4 g} \sigma_1^{n_1} \sigma_2^{n_2}$ with $n_1+ n_2= 6 g$ ($g$ is genus). Thus we first perform Taylor expansions of $c'$, $\sigma_1$ and $\sigma_2$ for each fixed genus $g$. The contour integral around $y_1=0$ and $y_2=0$ depends on the order of $y_i$ (i=1,2). For instance, genus one case of $n_1=2,n_2=4$, the integral of $y_2$ is firstly done, and secondly $y_1$ integral is evaluated, for a
non vanishing result. Opposite order gives a null result. Thus the contour integral of $y_i$ is non-commutative. There are  poles at $y_1= \pm y_2$  in the contour integral for $c'^4 \sigma_1^2 \sigma_2^4$ term.
Their contributions are cancelled, however. After taking all residues, we have 
\ba
U(\sigma_1,\sigma_2) &= & 4 \biggl(  \frac{3^3}{4} c'^4 ( \sigma_1^2 \sigma_2^4 + \sigma_1^4\sigma_2^2) - \frac{3^5}{40} c'^8 \sigma_1^6 \sigma_2^6
  + \frac{3^5}{160}c'^8(\sigma_1^2 \sigma_2^{10} + \sigma_1^{10}\sigma_2^2)\nonumber\\
 &&+O(c'^{16})
\biggr), 
\ea
where the residues of the term of $c'^8 \sigma_1^4\sigma_2^8$ for $y_1=0$ and $y_2=0$ have opposite signs when the order of evaluation of $y_i$ is changed, then sum of two terms is  cancelled. The term
of $c'^8 \sigma_1^2\sigma_2^{10}$ has a non-vanishing value by the order of the first $y_2=0$ residue, and the second $y_1=0$ residue. Opposite order of residue gives vanishing residue. 

From the selection rule  (\ref{6g}), the term of order of $c'^4$  corresponds to genus one, and the term of order of $c'^8$ to genus two.
This  confirms that  the two-point function is described by R-punctures ($l_1=l_2=-1$) and satisfies the selection rule (\ref{6g}). The values of $n_1$ and $n_2$ are even integers, and
there is no contribution from odd integer $n_i$ in two point function $U(\sigma_1,\sigma_2)$.

\vskip 2mm
{\bf (ii-b) p = $\frac{3}{2}$}
\vskip 2mm
The selection rule becomes from (\ref{RR}) for $p= \frac{3}{2}$,
\be
\frac{10}{3} g = n_1+n_2
\ee
when $l_1= l_2= -1$ (R type).
The genus $g$ is an integer, which leads to $g= 3 m$ ($m=1,2,3,...$). The factor of the exponent of (\ref{p1/2 int}) becomes for $p=\frac{3}{2}$ ($i=1,2)$,
\be
g(\sigma_i^{-\frac{1}{2}}y_i) = e^{\sum_i c' \sigma_i (5 y_i^3 - \frac{10 \sigma_i^2}{y_i} + \frac{\sigma_i^4}{y_i^5})}
\ee
and the other terms are the same as in the expression (\ref{p1/2 int}).
From the residues of $y_1=0, y_2=0$, the two-point function for the R type ($l_1=l_2=-1$) is expanded in the series of $c'^{4g/3} \sigma_1^{n_1}\sigma_2^{n_2}$ with 
$g= 3m$, $n_1+n_2= 10 m$ ($m$=1,2,...),
\ba
&&U_R(\sigma_1,\sigma_2) =  4 \biggl( - \frac{5^3}{4} c'^4 ( \sigma_1^2 \sigma_2^8 + \sigma_1^8\sigma_2^2) - 5^3 (\sigma_1^3 \sigma_2^7 + \sigma_1^7 \sigma_2^3)
\nonumber\\
&& - \frac{5^3 \cdot 3}{2} (\sigma_1^4 \sigma_2^6+\sigma_1^6 \sigma_2^4)
 + O(c'^8)\biggr), \hskip 5mm (p = \frac{3}{2})
\ea
The term of $c'^4 \sigma_1^5 \sigma_2^5$ has vanishing residue at $y_1=y_2=0$ and $y_1= \pm y_2$.
This expansion  is consistent with the selection rule  for $g=3m,(m=1,2,3,...)$ and 
$\frac{10}{3}g = n_1+n_2$ with $c'^{4g/3}$.
\vskip 2mm
{\bf (ii-c) p = - $\frac{1}{2}$   }
\vskip 2mm

For $p=-\frac{1}{2}$, we have
\be
g(\sigma^{-\frac{1}{2}}y) = e^{c'\sigma \frac{1}{y}}
\ee
Using the same integral for $U(\sigma_1,\sigma_2)$ in terms of the $g(y)$, we find from the residues at $y_1=0$ and $y_2=0$ 
\be\label{zeroU}
U(\sigma_1,\sigma_2) = 0
\ee
In this case, we note that the  selection rule is somewhat strange .
Indeed the equality  $3g -3 + s= \sum n_i + (g-1)(1 - \frac{2}{p}) + \sum l_i$, becomes
for $p=-\frac{1}{2}, s=2$,
\be
- 2g +4 =n_1+n_2 - 2 (l_1+l_2)
\ee
The $\sigma$ dependence is $\sigma^{(2g-1)(1+ \frac{1}{2})}=\sigma^{-(2g-1)}$. The expansion in powers of $\sigma$  assumed that the power $-(2g-1)$
was positive which is not satisfied for positive $g$. This strange situation occurs only for $p=-\frac{1}{2}$. The vanishing
result (\ref{zeroU}) for the two-point correlation function $U(\sigma_1,\sigma_2)$   may correspond to this strange selection rule.
For the other
cases, the two-point function is computed as in the case of the R-puncture for $p=\frac{1}{2}$ , and  the selection rule  is satisfied.

 \vskip3mm
 \section{Random supermatrices and duality}
  
 We first briefly summarize the results of our previous  study of random supermatrices, since it is needed for the present  discussion. We refer for  more details to \cite{BrezinHikami3}.
 
 A supermatrix $M$ is of the form
   \be\label{eq1}
   M = \left(\begin{array}{cc}
a & \alpha \\\bar{\alpha}  & b
\end{array}\right)\ee
where $a$ and $b$ are $n\times n$ and $m\times m$ Hermitian matrices, respectively. The rectangular matrices $\alpha$ and $\bar \alpha$ are $n\times m$ and $m\times n$  respectively, and their elements are Grassmannian
(i.e. anticommuting) variables.
 
   The average of  characteristic polynomials of the random supermatrix $M$,  in the presence of an external source $A$,  is defined  by
   \ba\label{duality1}
  F_k (\lambda_1,...,\lambda_k) &=& \frac{1}{Z_N}< \prod_{\alpha=1}^k \frac{1}{{\rm sdet}(\lambda_\alpha\cdot I - M)} >_{A,M}
  \nonumber\\
  &=& \frac{1}{Z_N}\int dM \prod_{\alpha=1}^k \frac{1}{{\rm sdet}(\lambda_\alpha\cdot I - M)}e^{\frac{i}{2}{\rm str} M^2 + i{\rm str} M A}
  \ea
  where $M$ and $A$ are supermatrices of the type (\ref{eq1}),  $I$ is the identity matrix and $Z_N$ is the normalization constant of the probability measure for $A=0$ ; the notations ${\rm str}$  abd ${\rm sdet}$ stand for supertrace and  superdeterminant   \cite{BrezinHikami3}.
We have to deal  with a complex weight to make meaning of the integrals since 
\be s\tr M^2 = \tr (a^2) -\tr (b^2)+ 2 \tr {(\al {\overline \al} )}\ee
  
  There is again a  duality formula for  (\ref{duality1})  \cite{BrezinHikami3} which gives the same $F_k$ of (\ref{duality1}) by  another average 
  \be\label{duality2}
F_k (\lambda_1,...,\lambda_k) = e^{i\sum_{1}^k \lambda_a^2/2}\int dB e^{\frac{i}{2} {\rm tr}B^2 + i \sum_{a=1}^k \lambda_a B_{aa}}\frac{\prod_{j=1}^m {\rm det}(B -\rho_j)}{\prod_{i=1}^n {\rm det}(B - r_i)} 
\ee
where the external source $A$ is given by $A = {\rm diag} (r_1,...,r_n,\rho_1,...,\rho_m)$ and  $B$ is a Hermitian $k\times k$ matrix.

 This is again an $N$-$k$ duality, which exchanges  the roles of the size $N=n+m$  of the matrices and of  $k$ the number of points,  between  $(\ref{duality1})$   and $(\ref{duality2})$. The  large $N$ scaling limit for the expectation values of the super-characteristic polynomials may then be  approached  from the dual representation  $ (\ref{duality2})$ after tuning of the external source matrix $A$.

Note that if  the "lower" eigenvalues $\rho_j=0$ (j=1,...,m) of the external source  vanish,  we obtain a simple determinant from (\ref{duality2}) 
\be\label{dual}
F_k (\lambda_1,...,\lambda_k) = e^{i\sum_{1}^k \lambda_a^2/2}\int dB ({\rm det} B)^m  e^{\frac{i}{2} {\rm Tr}B^2 + i \sum_{a=1}^k \lambda_a B_{aa} - \sum_1^n {\rm Tr } {\rm log}( B - r_i)}
\ee

Following the same strategy   we expand  ${\rm  log}(B- r_i)$ in  powers of $1/r_i$ : 
  
  \be
 \sum_{i=1}^n log (1- B/r_i) = - \sum_{i=1}^n \frac{1}{r_i} B - \sum_{i=1}^n \frac{1}{2 r_i^2} B^2 - \sum_{i=1}^n \frac{1}{3 r_i^3} B^3 - \cdots
  \ee
  Restricting now to sources  fulfilling  the conditions
  \be\label{condition2}
  \sum_{i=1}^n \frac{1}{r_i} = 0,\hskip 3mm \sum_{i=1}^n \frac{1}{r_i^2}= i n, \hskip 3mm \sum_{i=1}^n \frac{1}{r_i^3}= i n
  \ee
  we obtain from (\ref{dual}) the Kontsevich-Penner \cite{BrezinHikami30}. 
    \be\label{partition}
  F_k(\lambda_1,...,\lambda_k) = \frac{1}{Z_k}\int dB e^{i\frac{n}{3}{\rm Tr} B^3 + m {\rm Tr}{\rm log B} + i {\rm Tr} B\Lambda}
  \ee
  where  $B$ is a $k\times k$ Hermitian matrix, and $\Lambda = {\rm diag}(\lambda_1,...,\lambda_k)$.
 Indeed after the  rescaling   $B\to B/n^{1/3}$ the powers  $B^l$  with $l>3$ vanish in the large $n$ scaling limit.
  Thus we have obtained  the Kontsevich-Penner model from the supermatrix duality of (\ref{duality1}) and (\ref{duality2}). The supermatrix formulation provides  a natural derivation of the logarithmic term of the Penner model. Note that the Kontsevich-Penner model (\ref{partition}) had been
 obtained earlier  from a two-matrix model (or equivalently to a time dependent matrix model) \cite{BrezinHikami03}.
  
  As discussed in a previous article \cite{BrezinHikami03}, by adjusting  the constraints  (\ref{condition2}), we may obtain a generalized Airy matrix model with a
  logarithmic potential
    \be\label{partition2}
  F_k(\lambda_1,...,\lambda_k) = \frac{1}{Z_k}\int dB e^{\frac{c}{p+1}{\rm Tr} B^{p+1} + m {\rm Tr}{\rm log B} + i {\rm Tr} B\Lambda}
  \ee
  where $c$ is a constant given by the constraints on the source matrix.

     If  now, instead of letting all the  $\rho_j$ eigenvalues of the source vanish, we  perform also an expansion of the 
   numerator $\prod_{j=1}^m {\rm det}(B -\rho_j)$  in the integrand of $(\ref{duality2})$, we obtain
   \ba
  && ({\rm det}B)^m exp [{\sum_{j =1}^m {\rm Tr}{\rm log}(1 - \frac{\rho_j}{B})}]\nonumber\\
  &=& ({\rm det}B)^m exp \left(-{\rm Tr} [\frac{1}{B} \sum_{j=1}^m \rho_j  + \frac{1}{ 2 B^2}\sum_{j=1}^m \rho_j^2 + \frac{1}{3 B^3}\sum_{j=1}^m \rho_j^3 + \cdots]\right)
   \ea
   This yields a generalized Kontsevich-Penner model  with positive and negative powers of $B$,
   \be\label{c}
   F_k(\lambda_1,...,\lambda_k) = \frac{1}{Z_k}\int dB \ {\rm exp} [{ \frac{c}{p+1}{\rm Tr} B^{p+1} + \sum_l { c_l}{\rm Tr }\frac{1}{B^{l}}+ m {\rm Tr}{\rm log B} + i {\rm Tr} B\Lambda}]
  \ee
where  the scaling limit is here a large-$m$  limit, with $c$ a constant, and  $c_l=\frac{1}{l}\sum \rho_j^l$. 

  We now consider the $s$-point function defined as
  \be\label{correlation}
  U(t_1, ..., t_s) = < {\rm str} e^{i t_1 M} \cdots  {\rm str} e^{i t_s M}>
  \ee
  where  $t_i = -i \sigma_i$ $(i=1,2,...,s)$ are real parameters, ${\rm str}$ is the supertrace, and $M$ the supermatrix defined in (\ref{duality1}). This correlation function is a generating function for the intersection numbers when the external source is chosen at prescribed critical values.
  
  This correlation function $U(t_1, ..., t_s)$ may be related to (\ref{duality1}).  Indeed replacing 
  \be \prod_{\alpha=1}^k \frac{1}{{\rm sdet}(\lambda_\alpha\cdot I - M)}
  = e^{-str \sum_\alpha{\rm log}(\lambda_\alpha\cdot I - M)}.
  \ee
 and taking all the $\lambda_\alpha = \lambda$ equal, we have for one marked point,
  \ba
  U(\sigma) &=& \lim_{k\to 0} \frac{1}{k}\int d\lambda e^{\sigma \lambda} \frac{\partial}{\partial \lambda} < e^{k {\rm str} {\rm log} (\lambda - M)}>\nonumber\\
  &=& \int d\lambda e^{\sigma \lambda}<{\rm str} \frac{1}{\lambda - M}> = \int d\lambda e^{\sigma \lambda}<{\rm str} \delta(\lambda - M)> \nonumber\\
  &=& <{\rm str} e^{\sigma M}>
  \ea
  This one-point function is the Fourier transform of the resolvent.
  In  \cite{BrezinHikami3} we have established explicit expressions for the correlation functions  (\ref{correlation}) as contour integrals.  For the one point function ii reads
    \be\label{u(t)}
  U(\sigma) = \frac{1}{\sigma}\oint \frac{du}{2i\pi}e^{-i\sigma u}  \prod_{i=1}^n \frac{u - r_i + \sigma/2}{u - r_i - \sigma/2}
   \prod_{j=1}^m \frac{u - \rho_j - \sigma/2}{u - \rho_j + \sigma/2}
  \ee
For  $m=0$, it reduces to the ordinary  non-supersymmetric expression with external source $r_i$. The integration in the $u$-plane encircles  all the poles of the integrand (see  \cite{BrezinHikami3}). 

Similarly the connected part of the two point function $U_c(\sigma_1,\sigma_2)$  for  the supersymmetric case is,
  \ba\label{super2}
   U_c(\sigma_1,\sigma_2)  &=&  \oint \frac{du_1}{2i\pi} \frac{du_2}{2i\pi}e^{-i\sigma_1u_1-i\sigma_2u_2} \prod_1^n \frac {(u_1 -r_i +\sigma_1/2)(u_2-r_i+\sigma_2/2)}{(u_1 -r_i -\sigma_1/2)(u_2-r_i-\sigma_2/2)}\nonumber\\
&&\times \prod_1^m \frac{ (u_1 -\rho_j-\sigma_1/2)(u_2-\rho_i-\sigma_2/2)}{(u_1 -\rho_j +\sigma_1/2)(u_2-\rho_j +\sigma_2/2)}\nonumber\\
&&\times \frac {1}{ (u_1-u_2 -\sigma_1/2-\sigma_2/2)(u_1-u_2 +\sigma_1/2+\sigma_2/2)} \ea

  It generalizes the non supersymmetric case by the inclusion of $r_i$ and $\rho_j$  in the external source which provides an additional freedom.   To discuss the Airy matrix model with a logarithmic potential (Kontsevich-Penner model),  we  have chosen the simplest external sources $r_i=1$ and $\rho_j=0$.

An interesting application of the above formulae concerns the admixture of R and NS-punctures in the two-point function $U(\sigma_1,\sigma_2)$.  One can mix the case of $p=2$  (Kontsevich model) with a $p=\frac{1}{2}$ spin-curve in (\ref{super2}). As seen earlier the  $p=\frac{1}{2}$, is of R-type (fermionic
 vertex insertion) wheras  $p=2$ corresponds to an NS-puncture (bosonic vertex insertion).

  The scattering amplitudes for  superstring theory is related to the super Riemann surfaces with R and NS-punctures. Our present approach with super random matrices in an external source, provides the intersection numbers for punctures of both  types at arbitrary genus.   
 For  two different spins $p$ and $p'$,  the $n$-point function,  with a source choice similar to (\ref{pspin}) for $r_i$ and $\rho_j$, reads
 \ba\label{ppoint}
&&U(\sigma_1,....,\sigma_n) = <{\rm tr}e^{\sigma_1 M} \cdots {\rm tr}e^{\sigma_n M} >\nonumber\\
&=& \oint \prod_{i=1}^n \frac{du_i}{2i \pi} e^{c_1 \sum_{i=1}^n [(u_i+\frac{\sigma_i}{2})^{p+1}-(u_i - \frac{\sigma_i}{2})^{p+1}]}
 e^{c_2 \sum_{i=1}^n [(u_i+\frac{\sigma_i}{2})^{p'+1}-(u_i - \frac{\sigma_i}{2})^{p'+1}]}\nonumber\\
&\times& {\rm det}\frac{1}{u_i-u_j + \frac{1}{2}(\sigma_i+\sigma_j)}
\ea
with $c_1= \frac{n}{p^2-1}\sum_{i=1}^{p-1} \frac{1}{r_i^{p+1}}$, and  $c_2= \frac{m}{p'^2-1}\sum_{i=1}^{p'-1} \frac{1}{\rho_i^{p'+1}}$. 

The parameters  $c_1$ and $c_2$ are associated with  the spin $p$ and $p'$, and they are useful for  distinguishing the two different types,
for instance for the case when $p$ is of NS- type and $p'$ of R-type. An example is $p=2$ and $p'=\frac{1}{2}$, since they belong to 
NS and R-types, respectively, as we have seen in Section 3.

We investigate the admixture of  NS and R-punctures with $p=2$ and $p'=\frac{1}{2}$.
\ba\label{2ppoint}
&&U_c(\sigma_1,\sigma_2) = <{\rm tr}e^{\sigma_1 M}  {\rm tr}e^{\sigma_2 M} >_c \nonumber\\
&=& -\oint \prod_{i=1}^2 \frac{du_i}{2i \pi} e^{c_1 \sum_{i=1}^2 [(u_i+\frac{\sigma_i}{2})^{p+1}-(u_i - \frac{\sigma_i}{2})^{p+1}]}
 e^{c_2 \sum_{i=1}^2 [(u_i+\frac{\sigma_i}{2})^{p'+1}-(u_i - \frac{\sigma_i}{2})^{p'+1}]}\nonumber\\
&\times& \frac{1}{(u_1-u_2 + \frac{1}{2}(\sigma_i+\sigma_j))(u_2-u_1 + \frac{1}{2}(\sigma_i+\sigma_j))}
\ea
The selection rule is ,
\be
3 g -3 + s = n_1+n_2 + (g-1)(1-\frac{2}{\tilde p}) + \frac{1}{\tilde p}(l_1+l_2)
\ee
in which  $\tilde p$ means either $p$ or $p'$ and here $s=2$. 

For $p=2$, we take $l=0$ as NS-puncture, and for $p'=\frac{1}{2}$, the puncture is of R-type with $l=-1$.
We take $l_1=0$, and $l_2=-1$. 
For $p=2$, the factor $(g-1)(1-\frac{2}{p})$ vanishes, while for $p'=\frac{1}{2}$ , it is equal to $-3(g-1)$.
Therefore, the above selection rule  reads
\be
6g -6 + s = n_1 + n_2 -2
\ee
with $U_c(\sigma_1,\sigma_2) \sim \sum \sigma_1^{n_1+ \frac{1}{2}}\sigma_2^{n_2}$, with punctures of NS-type ($\sigma_1$) and R ($\sigma_2$), respectively.

 We have discussed  negative values of $p$ for the matrix model of section 3. 
Matrix models with a logarithmic potential and negative powers of the matrices have been discussed in connection with
 superconformal gauge fields in the irregular conformal limit  \cite{Gaiotto,NishinakaRim}.

   \vskip 2mm 
\section{Open Intersection numbers }

\vskip2mm

As we have discussed in the previous section, an Airy matrix model with a logarithmic potential (Kontsevich-Penner model) may be  derived from (\ref{partition})
 and its intersection numbers
are deduced from the generating functions $U(\sigma_1,...,\sigma_n)$ 
by an appropriate tuning of the external matrix source. 

For $p=2$ the one point function $U(\sigma)$ with a boundary reads 
\ba\label{integral}
U(\sigma) &=& \frac{1}{\sigma} \oint \frac{du}{2i\pi} e^{-\frac{c}{3}[(u + \frac{\sigma}{2})^3 - (u - \frac{\sigma}{2})^3] + m {\rm log}(u +\frac{\sigma}{2})-m{\rm log}(u - \frac{\sigma}{2})}\nonumber\\
&=& \frac{1}{\sigma}e^{- \frac{c}{12} \sigma^3}\oint \frac{du}{2i\pi} e^{-c \sigma u^2 + m {\rm log}(u + \frac{1}{2}\sigma)-m{\rm log}(u-\frac{1}{2}\sigma)}
\ea
 The constant $c$ is related to the normalization of the eigenvalues of the external source as in (\ref{c}). Let us denote
\be
\sigma = \frac{1}{\lambda}, \hskip 3mm t_n= \frac{1}{\lambda^{n+\frac{1}{2}}}.
\ee
For $m=0$, we have
\be
U(\sigma) = \sqrt{\frac{\pi}{c}}\sum_{g=1}^\infty \frac{(-c)^g}{(12)^g g!}t_{3g-2}
\ee
From the above expression, we deduce the intersection number $<\tau_{3g-2}>$,
\be
<\tau_n>= \frac{1}{(24)^g g!}
\ee

where  $n$ is given by Riemann-Roch formula (which will be discussed below in (\ref{Riemann})), as $n= 3g- 2$.

We note here the  low orders for the intersection numbers for $m \neq 0$, computed  from the integral  (\ref{integral}),
\ba\label{tau}
<\tau_{1}> &=&\frac{1+12 m^2}{24},\hskip 2mm <\tau_{\frac{5}{2}}> = \frac{m+ m^3}{12},\hskip 2mm <\tau_4> = \frac{1 + 56 m^2+ 16 m^4}{1152} \nonumber\\
<\tau_{\frac{11}{2}}> &=& \frac{12m + 25 m^3+ 3 m^5}{2880}, ...
\ea

These numbers have been   obtained earlier from the  Virasoro equations, within the replica method in \cite{BrezinHikami4}, and the results of  both methods agree.The  above results coincide also with those of  \cite{Bertola}, if we replace $m$ by their parameter N, which is the size of the matrices. 
The indices $n$ of $\tau_n$ in (\ref{tau}) are integers or half-integers. If $n$ is a half-integer, the marked point is considered to be located on the boundaries.

The one point intersection numbers (\ref{tau}) are easily obtained from $U(\sigma)$ for $m\neq 0$. After rescaling of $u$, one has
\be\label{U}
U(\sigma) = \frac{e^{-\frac{c}{12}\sigma^3}}{2\sigma^{3/2}}\oint \frac{du}{2i\pi}e^{-\frac{c}{4}u^2+ m{\rm log}(\frac{u+\sigma^{3/2}}{u- \sigma^{3/2}})}
\ee
The logarithmic term is expanded in powers of $\sigma^{3/2}$ 
\be\label{logexpand}
{\rm log}(\frac{u + \sigma^{3/2}}{u-\sigma^{3/2}}) = \frac{2}{u}\sigma^{3/2}+\frac{2}{3u^3}\sigma^{9/2} +\frac{2}{5 u^5}\sigma^{15/2} +\cdots
\ee
For odd powers of $u$ in the integrand of (\ref{U}) , i.e. $u^{-(2j+1)}$, the integral is  just a contour integral around $u=0$.
\be\label{contoura}
\oint \frac{du}{2 i \pi} \frac{e^{-a u^2}}{u^{2k+1}} = \frac{(-a)^k}{k!}
\ee
 For even powers of $u$, i.e. $u^{-2j}$ in the integrand, the integration becomes non-local. 
 The following integral $I$ leads to $\Gamma$- functions,
\ba\label{Gamma}
I &=& \int_{-\infty}^\infty du e^{-a u^2}\frac{1}{u^{2k}}=\int_0^\infty dt t \hskip 1mm t^{-\frac{1}{2}-k} e^{-at}\nonumber\\
&=& a^{k-\frac{1}{2}}\Gamma(\frac{1}{2}- k) = (-1)^k \frac{2^k\sqrt{\pi}}{(2k-1)!!}a^{k-\frac{1}{2}}
\ea
in which a continuation from positive to negative  $k$ has been used. 

For instance, for even powers of $m$, we have up to order $m^2 \sigma^3$,
\ba
U(\sigma) &=& e^{-\frac{c \sigma^3}{12}}\frac{1}{2\sigma^{3/2}}\int e^{-\frac{c}{4}u^2}(1 + 2 m^2 \sigma^3 \frac{1}{u^2})
\nonumber\\
&=& \frac{1}{2\pi}\sqrt{\frac{\pi}{c}}e^{-\frac{c\sigma^3}{12}}\frac{1}{\sigma^{3/2}}(1 - c m^2\sigma^3)
\ea
which indeed reproduces the result   (\ref{tau}),
\be\label{tau1}
<\tau_1> =\frac{1}{24}(1 + 12 m^2)
\ee
 Using the formula of (\ref{contoura}) and (\ref{Gamma}), we have  derived the one-point intersection numbers  to all orders in the genus, and they agree with the results derived by other methods \cite{BrezinHikami4, Bertola}. 
 
 In the appendix we recall how to use the Virasoro constraints for a matrix model with a logarithmic potential. This leads to explicit results for the intersection with NS and R punctures.

\vskip 2mm
{\bf Number of boundary components  for the Kontsevich-Penner model}
\vskip 2mm
The parameter $m$ of the Kontsevich-Penner model, ($m$ is the coefficient of the logarithmic potential), is related to the number of boundaries as follows.  The number $b$ of tboundary components appears as  the power of $m$, i.e. $m^b$
\cite{Safnuk,Bertola}.
To convince oneself of the validity of this interpretation one considers the expansion
 \be
e^{m {\rm tr log} M } = ({\rm det} M)^m = \sum _{b=0}^\infty \frac{1}{b!} m^b ({\rm tr log} M)^b
\ee
Thus the correlation function $U(s)$ has $b$-boundaries, described by  the insertion of  ${\rm tr log} M= {\rm log det} M$.
In general, we could consider the moduli space with genus $g$, n
$b$ boundaries, $n$ interior marked point, and $k$ marked points at the boundary. This moduli space is denoted as $M_{g,b;k,n}$.

Since the Riemann surface with boundary becomes a Klein surface, we interpret it as a double surface $D\Sigma$ (D means double), where $\Sigma$ has genus $g$, with $b$ boundary components and $n$ interior marked points.
The surface $D\Sigma$ has a doubled genus $g= 2h+ b-1$ , where $h$ is the number of handles and $b$ the number of boundary components. We have assumed here that there are no marked points at the boundary.
From the dimensional constraint, the Riemann-Roch theorem gives
\be\label{Riemann}
3g-3+ s = 6 h  - 6 + 3b + 2n = 2 \sum_{i=1}^n n_i
\ee
where $n$ is the number of interior marked points (double counting; s=2n) and $n_i$ the indices of the intersection numbers $<
\prod_{i=1}^n \tau_{n_i}>$, 
$b$ is the number of boundaries (holes) and  $n$ is the number of punctures. 
 
 For the one point correlation,  we have seen that when the intersection numbers are expressed as  polynomials in $m$,  when the power of $m$ is odd, the coefficient is obtained by a residue calculation, and it leads to an R-type. If the power of $m$ is even, then
 the puncture is of NS-type.

 The interpretation of the Kontsevich-Penner model for the moduli space with  boundaries does not  refer to  marked points at the boundaries.
 It has been argued  that when $m=1$, it describes the moduli space  with marked point at the boundaries \cite{Alexandrov}.
 In any case,  it does not answer the question about how these marked points are distributed on the boundaries. To answer this question of the distribution of  marked points, 
 a refined open Kontsevich-Penner model has been  proposed \cite{AlexandrovBuryak}.

   \vskip 3mm
  {\section{ Summary}}
\vskip 3mm
  In this article we have considered matrix models with a $p$-spin structure which generalize Kontsevich Airy matrix model.   The external source plus duality method that we have used, provides explicit integral representations for the generating functions of the intersection numbers. Therefore these matrix models provide an alternative approach to the computation of the intersection numbers. 
  
   The integral representations of the generating functions  present $p$ Stokes domains which characterize the so-called spin-structure in our formulation of the problem. The value $l=-1$ of the spin component corresponds to a Ramond puncture, but such punctures are not present for integer $p$. However our formulation allows for  a continuation  to non-integer and negative  values of $p$.  We could  then show that the matrix models for half-integer spins $p$  do present  R-type punctures.
  The results that we have found for the intersection numbers  confirm the selection rule given by the Riemann-Roch relation, and allow for an extension of this rule  to half-integer spins.
  
  We have also used integrals over super-matrices that we had introduced in an earlier work,  within the same framework of external source and explicit correlation functions. The scaling limit leads then to a natural extension to matrix models with logarithmic potentials which are known to generate surfaces with boundaries which are also considered in this work.
  
  We intend to study future extensions to more general symmetries.

 \vskip 3mm

 {\bf Acknowledgement}
 \vskip 3mm
 We are thankful to  Edward Witten for a  correspondence about Ramond punctures.   S.H.  is partially supported by JSPS KAKENHI, 19H01813. E.B. thanks  Okinawa Institute of Science and Technology (OIST) for its kind hospitality.
  
 \vskip 6mm
  {\bf {Appendix : intersection numbers for logarithmic potentials}}
  \vskip 3mm
      
 With the logarithmic potential introduced in \cite{BrezinHikami4}, we have for $p=2$ (Kontsevich model),
 \be
 Z = \int dB e^{{\rm tr}( - \frac{1}{3}B^{3} + B \Lambda + m {\rm log} B)}
 \ee
 and $F = {\rm log} Z$ has now additional terms $\Delta F$  for $m\ne 0$ as in \cite{BrezinHikami4},
 \be\label{Ramondpair}
 \Delta F = \frac{1}{2} m t_0 t_{\frac{1}{2}} + \frac{1}{4}m^2 t_1 + \frac{1}{16}m^2 t_1^2 + \frac{m}{4}t_0t_{\frac{1}{2}}t_1+ \frac{1}{24}m t_{\frac{1}{2}}^3 + \frac{1}{4}m^2 t_{\frac{1}{2}} t_{\frac{3}{2}} + \cdots
 \ee
 Note that the  $t_{\frac{1}{2}}$ can be written as ($t_{1,-1}$) according to the previous notation of $t_{n,l}$, which is $\sum \frac{1}{\lambda_i^{n+\frac{1}{p}(l+1)}}$. It belongs to  an R-puncture since it has $l=-1$.  The Virasoro equations follow   from the constraints
 \be
 \int dB \frac{\partial}{\partial B_{ba}} e^{{\rm tr}(-\frac{1}{3}B^3 + B \Lambda + m {\rm log} B)} = 0
 \ee
 i.e.
 \be
 (-(\frac{\partial}{\partial \Lambda})^3_{ab} + (\Lambda^T \frac{\partial}{\partial \Lambda})_{ab} + (k + m)\delta_{ab}) Z = 0
 \ee
The partition function $Z$ may then be obtained  for finite $k$ ($k$ is a size of the matrix $B$),  as an expansion in inverse powers of the  $\lambda_i$ \cite{BrezinHikami4}.
 
 There are no R-punctures in the Kontsevich model without logarithmic potential,   but R-pairs do  appear in the presence of a logarithmic potential,  in the case of even
 powers of $m$, as seen in the last term of (\ref{Ramondpair})  $\frac{1}{4}m^2 t_{\frac{1}{2}}t_{\frac{3}{2}}= \frac{1}{4}m^2 t_{1,-1}t_{2,-1}$.
 The power of $m$ gives  the number of boundaries. The existence of  a Ramond sector for a logarithmic potential
 had already been noticed in \cite{BrezinHikami4}.
 
 When $p=2$ with logarithmic potential, the string equation is
 \be
 (- \frac{\partial}{\partial t_{0,0}} + \frac{1}{4}t_{0,0}^2 -\frac{m}{2}t_{1,-1}+ \sum_{n=0,\frac{1}{2},1,\frac{3}{2}, 2,...} (n+ \frac{1}{2}) t_{n+1,0}\frac{\partial}{\partial t_{n,0}}) g = 0
 \ee
 where $Z=Z_0 g$ with 
 \be
 Z_0 = \frac{1}{\prod_{i<j} (\sqrt{\lambda_i}+\sqrt{\lambda_j})^{\frac{1}{2}}}e^{\frac{2}{3}\sum \lambda_i^{\frac{3}{2}}}
 \prod \lambda_i^{\frac{m}{2}}
 \ee
 The string equation leads to
 \be
 <\tau_{0,0}\prod_i \tau_{n_i,l_i}> = \sum_i <\tau_{n_i-1,l_i}\prod_{j\ne i} \tau_{n_i,l_i}>
 \ee
 In the presence of a logarithmic potential, there is an additional Virasoro equation for $t_{1,-1}$.
 \be
 (-2 \frac{\partial}{\partial t_{1,-1}} - m t_{0,0} - (\frac{1}{16}+  \frac{m^2}{4})t_{2,-1} - \frac{1}{12}J_{-4}^{(3)} + \frac{m}{4}J_{-4}^{(2)}
 -\frac{1}{2}J_{-1}^{(2)} ) g = 0
 \ee
 where we  have used the notations of $J_{-m}^{(l)}$ from \cite{BrezinHikami4}. The intersection numbers are tabulated in \cite{Bertola}.
  
 The string equation works also for the R-punctures. For instance, we have the relation,
 \be
 <\tau_{0,0}\tau_{4,-1}^2> = 2 <\tau_{3,-1}\tau_{4,-1}> =\frac{1}{144}m^2(m^4+11m^2 + 16)
 \ee
The intersection numbers, with odd numbers  of R-punctures, such as $\tau_{n,-1}$,   may be expressed as polynomials
 in $m$,  with odd powers  of $m$.
 The intersection numbers, with an even number of R-punctures, are polynomials
 in $m$,  with even powers of  $m$. Hence, there is a parity for the numbers of R-punctures.

       \vskip5mm


\begin{thebibliography}{99}
\bibitem{'t} 
G. 't Hooft, A planar diagram theory for strong interactions, Nucl.Phys. B72 (1974) 461
\bibitem{Stanford}
D. Stanford and E. Witten, JT gravity and the ensembles of random matrix theory, arXiv:1907.03363.
\bibitem{BrezinHikami90} 
E. Brezin and S. Hikami, Universal singularity at the closure of a gap in a random matrix theory. Phys. Rev, E 57 (1998) 4140.
\bibitem{BrezinHikami01}
E. Br\'ezin and S. Hikami, Intersection theory from duality and replica, Communication in Mathematical Physics, 283 (2008), 507.
\bibitem{BrezinHikami02}
E. Br\'ezin and S. Hikami, Intersection numbers of Riemann surfaces from Gaussian matrix models, JHEP 10 (2007) 096.
\bibitem{BrezinHikami03}
E. Br\'ezin and S. Hikami, Computing topological invariants with one and two-matrix models, JHEP 04 (2009)110.
\bibitem{BrezinHikami5}
E. Br\'ezin and S. Hikami, Random matrix, singularities and open/close intersection numbers,  Journal of Physics A:Mathematical and Theoretical, 48 (2015) 475201.
\bibitem{BrezinHikami2}
   E. Br\'ezin and S. Hikami, Random matrix theory with an external source, SpringerBriefs in Mathematical Physics Vol. 19 (2016), Springer. 
\bibitem{Kontsevich}
M. Kontsevich, Intersection theory on the moduli space of curves and the matrix Airy function, Commun. Math. Phys. 147 (1992) 1.


   \bibitem{Semenov}
 A. Mironov, A. Morozov and G.W. Semenov, Unitary matrix integrals in the framework of generalized Kontsevich model 1. Brezin-Gross-Witten model,
Intern. J. Mod. Phys.  A, 11 (1996) 5031. arXiv:hep-th/9404005.
    \bibitem{Troost1}
 S. Li and J. Troost, Topological gravity with non-compact matter, JHEP01(2019)158.
   
 \bibitem{WittenP}
E. Witten, Algebraic geometry associated with matrix models of two dimensions, in "Topological methods in modern mathematics", Publish or Perish, INC.,(1993). P.235.
  \bibitem{Witten40}
   E. Witten, The N matrix model and gauged WZW models, Nucl.Phys. B371(1992)191.
   \bibitem{Harer}
 J. Harer and D. Zagier, The Euler characteristic of the moduli space of curves, Invent. Math. 85 (1986) 457.
 \bibitem{Penner}
R.C. Penner, Perturbative series and the moduli space of Riemann surfaces,  J. Diff. Geometry, 27 (1988) 35.
\bibitem{GrossWitten}
E. Witten and D. Gross, Possible third order phase transition in the large N lattice gauge theory, Phys. Rev. D 21 (1980) 446.
\bibitem{BrezinGross}
E. Br\'ezin and D. Gross, The external field problem in the large N limit of QCD. Phys. Lett. 97 (1980) 120.
\bibitem{Gross}
D.J.Gross and M.J.Newman, Unitary and Hermitian matrix models in an external field. 2: Kontsevich model and continuum Virasoro constraints, Nucl. Phys. B 380 (1992)168.

\bibitem{BrezinHikami70}
E. Br\'ezin and S. Hikami, Duality and replicas for a unitary matrix model, JHEP07 (2010) 067.
    \bibitem{BrezinHikami4}
E. Br\'ezin and S. Hikami, On an Airy matrix model with a logarithmic potential, J. Phys. A: Mathematical and Theoretical 45 (2012) 045203. 
\bibitem{Kac}
V.G. Kac and M. Wakimoto, 
Modular invariant representations of infinite-dimensional Lie algebras and superalgebras, Proc. Nat. Acad. Sci. USA 85 (1988) 4956.
\bibitem{Gaberdiel}
M.R. Gaberdiel, Fusion rules and logarithmic representation of a WZW model at fractional level, Nucl. Phys. B618 (2001) 407.
\bibitem{Lesage}
F. Lesage, P. Mathieu, J. Rasmussen and H. Saleur, The $\widehat{su}(2)_{-\frac{1}{2}}$ WZW model and $\beta \gamma$ systems,
Nucl. Phys. B647, 363 (2002), arXiv: 0207201.
\bibitem{Nekrasov}
 N.A. Nekrasov, Lectures on curved beta-gamma systems, pure spinors, and anomalies, (2005) arXiv:hep-th/9511008.
 \bibitem{Li}
 K. Li, Topological gravity with minimal matter, Nucl. Phys. B354 (1991) 711.
  \bibitem{Verlinde}
 E. Verlinde and H. Verlinde, Solution of two-dimensional topological quantum gravity, Nucl. Phys. B348 (1991) 457.
 \bibitem{Gomis}
J. Gomis, Z. Komargodski and N. Seiberg, Phases of adjoint QCD$_3$ and dualities, SciPost Phys. 5, 007 (2018).
 \bibitem{WittenR}
 E. Witten, The super period matrix with Ramond punctures, J. Geometry and Physics, 92 (2015) 210. arXiv:1501.02499.
 \bibitem{WittenS}
 E. Witten, Notes on super Riemann surfaces and their moduli, Pure and Applied Mathematics Quarterly 15 (2019) 57. arXiv:1209.2459.
\bibitem{Jarvis}
T. Jarvis, T. Kimura and A. Vaintrob, Moduli spaces of higher spin curves and integrable hierarchies. Composite Math. 126 (2001) 157.
\bibitem{Mochizuki}
T. Mochizuki, The virtual class of the moduli stack of stable $r$-spin curves, Commun. Math. Phys. 264 (2006) 1-40.
\bibitem{Fan}
 H. Fan, T. Jarvis and Y. Ruan, The Witten equation, mirror symmetry, and quantum singularity theory, Annals of Mathematics 178 (2013), 1-106.
\bibitem{Seeley}
R. Seeley and I.M.Singer, Extending $\bar \partial$ to singular Riemann surfaces, JGP. 5 (1981) 121.
\bibitem{Rarita}
W.Rarita and J. Schwinger, On a theory of particles with half-integral spin, Phys. Rev. 60 (1941) 61.
   \bibitem{BrezinHikami3}
 E. Br\'ezin and S. Hikami, Random supermatrices with an external source, JHEP 08 (2018) 086.   




 \bibitem{Gaiotto}
D.  Gaiotto, Asymptotically free N=2 theories and irregular conformal block, J. Phys: Conf. Ser. 462 (2013) 012014. arXiv:0908.0307.
   \bibitem{NishinakaRim}
 T. Nishinaka and C. Rim, Matrix models for irregular conformal blocks and Argyres-Douglas theories, JHEP10(2012)138.
\bibitem{BrezinHikami30}
E. Br\'ezin and S. Hikami, Characteristic polynomials of random matrices, Commun. Math. Phys. 214 (2000),111-135.


\bibitem{BrezinHikami66}
E. Br\'ezin and S. Hikami, The intersection numbers of the $p$-spin curves from random matrix theory,
JHEP02(2013)035.



\bibitem{Pandharipande}
R. Pandharipande, J. P. Solomon and R. J. Tessler, Intersection theory on moduli of disks, open KdV and Virasoro.  arXiv:1409.2191.
\bibitem{Alexandrov}
A. Alexandrov, Open intersection numbers, Kontsevich-Penner model and cut-and-joint operators. arXiv:1412.3772.
\bibitem{Safnuk}
B. Safnuk, Combinatorial models for moduli spaces of open Riemann surfaces, arXiv:1609.07226.
\bibitem{AlexandrovBuryak}
A. Alexandrov, A. Buryak and R. Tessler, Refinded open intersection numbers and the Kontsevich-Penner matrix models. JHEP 03 (2017) 123. arXiv:1702.02319.
\bibitem{Bertola}
M. Bertola and G. Ruzza, The Kontsevich-Penner matrix integral, isomonodromic tau functions and open intersection numbers. 
Ann. Henri Poincar\'e 20(2019)393-443. arXiv:1711.03360.


 \end{thebibliography}
 \end{document}